\documentclass[a4paper,10pt]{article}
\usepackage{amsmath}
\usepackage{amsfonts}
\allowdisplaybreaks[1]
\numberwithin{equation}{section}
\begin{document}
\baselineskip=7mm

\pagestyle{plain}

\hskip 12.36cm{}
\vspace{1cm}

\begin{flushright}
RUP-10-1  \,\,\,\,\,\,\,\,\,\,\,\,\, \\ 
\end{flushright}

\vspace{3cm}

\centerline{\bf \Large Low Energy Action of ``Covariant'' Superstring Field Theory}
\vskip 2mm
\centerline{\bf \Large in the NS-NS {\it pp}-Wave Background}

\vspace{1.5cm}
\setcounter{footnote}{0}
\renewcommand{\thefootnote}{\fnsymbol{footnote}}
\centerline{\bf Yoichi Chizaki$^{1,}$\!\!
\footnote{E-mail: {\tt y.chizaki@aist.go.jp}}
and Shigeaki Yahikozawa$^{2,}$\!\!
\footnote{
E-mail: {\tt yahiko@rikkyo.ac.jp}}}
\vspace{1mm}
\setcounter{footnote}{0}
\renewcommand{\thefootnote}{\arabic{footnote}}
\centerline{\it $^1$Nanosystem Research Institute, National Institute of Advanced Industrial Science }
\centerline{\it and Technology,
Tsukuba 305-8568, Japan}
\centerline{\it $^2$Department of Physics, Rikkyo University,
Tokyo 171-8501, Japan}

\vspace{1cm}

\centerline{\bf Abstract}
Exact construction of superstring field theory in some background fields is very important. We construct the low energy NS-NS sector of superstring field action in the pp-wave background with the flux of NS-NS antisymmetric tensor field (NS-NS pp-wave) without gauge fixing up to the second-order where the action is world-sheet BRST invariant. Here we use the word ``covariant'' in a invariant theory for a symmetric transformation of the pp-wave background which is not the Lorentz transformation in the flat background. Moreover we prove the exact correspondence between this low energy action and the second-order perturbation of supergravity action in the same background. We also prove the correspondence of the gauge transformation in both the actions. This construction is based on  
the BRST first quantization of superstrings in the pp-wave background in our previous paper.
\vspace{2mm}

\newpage
\section{Introduction}

Exact construction of superstring field theory in some general backgrounds is very important. Historically light-cone string field theory is firstly constructed in the flat background \cite{Kaku-Kikkawa1}\cite{Kaku-Kikkawa2}, after that, covariant string field theory is constructed \cite{Witten}-\cite{Zwiebach}. String field theory without gauge fixing is the theory that equations of motion and interactions are determined by the gauge symmetry so that the construction of string field theory in the general backgrounds enables us to obtain the deeper understanding of fundamental low of physics. At present in the case of the R-R pp-wave background, light-cone superstring field thory is constructed \cite{Spradlin-Volovich1}-\cite{Pankiewicz-Stefanski}, however complete ``covariant'' superstring field thory without gauge fixing in the background is not constructed yet. Recently exact first quantization of superparticle in the AdS background is constructed in \cite{Horigane-Kazama}, however first quantization of superstrings has not been constructed yet. As an example, the low energy state of superstring field theory in the flat background based on the first quantization is proposed by Berkovits \cite{Berkovits} so that we exactly prove the construction of the low energy state of superstring field theory in the NS-NS pp-wave background using the proporsal and results in our previous papers \cite{Chi-Ya1}\cite{Chi-Ya2}. 

As a first step for understanding superstring field theory in the general backgrounds, exact construction of superstring field theory in the case of pp-wave background has a deep meaning. The reason is that we can calculate superstring field theory in this background without contradiction and we can compare it with coresponding supergravity. Here the supergravity action is derived by comformal imvariance of superstring theory in some backgrounds \cite{Callan-Friedan-Martinec-Perry}.
As we know, taking the Penrose limit on some spacetimes near the null geodesic, they become a pp-wave background \cite{Penrose}, which is generalized by \cite{Gueven} and \cite{BFHP} in the case of the spacetime with NS-NS and R-R flux. So that first of all, it is important to perfectly understand superstring theory in this simple background before understanding some general backgrounds. Historically first quantization of (super)string theory in the pp-wave background have been performed by fixing the light-cone gauge for a long time \cite{Horowitz-Steif1}-\cite{Metsaev-Tseytlin} and after these research, the light-cone superstring field theory in R-R pp-wave background is constructed \cite{Spradlin-Volovich1}-\cite{Pankiewicz-Stefanski}, however we can not obtain deeper knowledge between superstring theory and background fields from the light-cone gauge quantization because it maps to a free field theory in this gauge fixing. Therefore we perform the BRST first quantization of superstring theory and construct the general solutions defined as Heisenberg operator solutions which are called the {\it general operator solutions} in the NS-NS pp-wave background in our previous papers \cite{Chi-Ya1}\cite{Chi-Ya2}.

 In this paper we exactly construct superstering field theory without interactions in the NS-NS pp-wave background based on the {\it general operator solutions} and free-mode expansions of them which are called the {\it free-mode representations} in our previous papers \cite{Chi-Ya1}\cite{Chi-Ya2}. Moreover, we particularly notice the low energy state of NS-NS sector of this string field whose construction is based on \cite{Berkovits}. Then we confirm the correspondence between NS-NS sector of supersteing field action and the second-order perturbation of supergravity action without R-R fields and fermionic fields in the same NS-NS pp-wave background. In this construction, the modes defined by the {\it general operator solutions} play an important roll, because these modes have the linear dependence of coordinates which enables us to reproduce the Christoffel symbol and the covariant derivative at the standpoint of superstring field theory. We also confirm the gauge symmetry in both supersting field theory and supergravity in this pp-wave background. 
  
This paper is organized as follows. In section 2, we prepare for the construction of superstring field theory in the NS-NS pp-wave background. This construction is based on the {\it general operator solutions} and the {\it free mode representations} that we construct in our previous paper so that first we write up them. We also explain the super-Virasoro operator and BRST operator in this background. In section 3, we define the low energy string field action. Here construction of the low energy state is important. In the construction we use new modes constructed from free modes. The caracteristic point of these modes is coordinate dependence which playes an important role. Because of the coordinate dependence we have to note the differential operator. In the last part of this seciton we write up the usefull formulae. In section 4, we calculate the component of superstring field action constructed in section 3. In section 5, we compare the low energy superstring field action to the second-order perturbation of supergravity action and we confirm the correspondence between them. In section 6, we compare the gauge transformation at the standpoint of superstring field theory to the corresponding supergravity. In section 7, we summerize this paper. In appendix A, we represent the BRST transformations of all the free modes in the NS-NS pp-wave background. In appendix B, we define the modes based on the {\it general operator solutions} and the {\it free-mode representations}.               


\vskip 1cm
\section{Preliminaries}
In this section we write up the definition based on our previous paper.
In our previous paper we constructed the general solutions and the covariant canonical quantization of the Ramond-Neveu-Schwarz (RNS) superstrings in the NS-NS pp-wave background. These geneal solutions and Heisenberg operator solutions are called the {\it general operator solutions}. The NS-NS pp-wave background is 
\begin{align}
ds^{2}&=g_{\mu\nu}dx^{\mu}dx^{\nu}=-{\mu^{2}z^{*}z}dx^{+}dx^{+}-2dx^{+}dx^{-}+dz^{*}dz
+dx^{k}dx^{k},\label{pp-wave:metric}\\
B_{+z}&=-\frac{i}{2}{\mu z^{*}},
B_{+z^{*}}=+\frac{i}{2}{\mu z},\label{NS-NS:flux}
\end{align}
where we use the light-cone coordinates $x^{\pm}=\frac{1}{\sqrt{2}}(x^{0}\pm x^{1})$, complex coordinates $z=x^{2}+ix^{3}$, $z^{*}=x^{2}-ix^{3}$, spacetime Greek indices run over $+,-,z,z^{*},k$ (where $k$ runs over $4,\cdots,9$) and the coefficient $\mu$ is a constant which denotes the strength of NS-NS pp-wave background. We also use the world-sheet light-cone coordinates $\sigma^{\pm}=\tau\pm\sigma$. First we explain the {\it general operator solutions} for bosonic fields and fermionic fields, however they tangle in each other complicatedly so that we explain the details after showing the forms of the solutions.
The {\it general operator solutions} for the world-sheet bosonic fields namely the string coordinates $X^{\mu}(\tau,\sigma)$ are 
\begin{align}
X^{+}(\tau,\sigma)&=X^{+}_{\rm L}(\sigma^{+})+X^{+}_{\rm R}(\sigma^{-}),\label{gos:+}\\
X^{-}(\tau,\sigma)&=X^{-}_{0\rm L}(\sigma^{+})+X^{-}_{0\rm R}(\sigma^{-})\nonumber\\
&\ \ \ +X^{-}_{1\rm B}(\sigma^{+},\sigma^{-})+X^{-}_{1\rm F}(\sigma^{+},\sigma^{-})
+\frac{i\mu}{2}\left[f(\sigma^{+})g^{*}(\sigma^{-})-f^{*}(\sigma^{+})g(\sigma^{-})\right],\label{gos:-}\\
Z(\tau,\sigma)&=e^{-i\mu\tilde{X}^{+}}
\left[f(\sigma^{+})+g(\sigma^{-})\right],\label{gos:z}\\
Z^{*}(\tau,\sigma)&=e^{i\mu\tilde{X}^{+}}
\left[f^{*}(\sigma^{+})+g^{*}(\sigma^{-})\right],\label{gos:z*}\\
X^{k}(\tau,\sigma)&=X^{k}_{\rm L}(\sigma^{+})+X^{k}_{\rm R}(\sigma^{-}).
\label{gos:k}
\end{align}
The {\it general operator solutions} for the world-sheet fermionic fields $\psi^{\mu}_{\pm}(\tau,\sigma)$ (where sub indeces $\pm$ denote spinor indeces) which are superpartner of the string coordinates are 
\begin{align}
\psi^{+}_{\pm}(\tau,\sigma)&=\psi^{+}_{\pm}(\sigma^{\pm}),\label{gos:psi+}\\
\psi^{-}_{\pm}(\tau,\sigma)&=\psi^{\ -}_{0\pm}(\sigma^{\pm})
\mp\frac{i\mu}{2}\left[f(\sigma^{+})+g(\sigma^{-})\right]
\lambda^{*}_{\pm}(\sigma^{\pm})
\pm\frac{i\mu}{2}\left[f^{*}(\sigma^{+})+g^{*}(\sigma^{-})\right]
\lambda_{\pm}(\sigma^{\pm}),\label{gos:psi-}\\
\psi^{Z}_{\pm}(\tau,\sigma)&=e^{-i\mu\tilde{X}^{+}}
\left[\lambda_{\pm}(\sigma^{\pm})\mp i\mu\psi^{+}_{\pm}(\sigma^{\pm})
\left\{f(\sigma^{+})+g(\sigma^{-})\right\}\right],\label{gos:psiz}\\
\psi^{Z*}_{\pm}(\tau,\sigma)&=e^{i\mu\tilde{X}^{+}}
\left[\lambda^{*}_{\pm}(\sigma^{\pm})\pm i\mu\psi^{+}_{\pm}(\sigma^{\pm})
\left\{f^{*}(\sigma^{+})+g^{*}(\sigma^{-})\right\}\right],\label{gos:psiz*}\\
\psi^{k}_{\pm}(\tau,\sigma)&=\psi^{k}_{\pm}(\sigma^{\pm}).\label{gos:psik}
\end{align}
We explain the details in the {\it general operator solutions}. L and R indicate the left-moving and the right-moving parts respectively whose fields are arbitrary bosonic periodic functions of $\sigma^{\pm}$. Here $\tilde{X}^{+}=X^{+}_{\rm L}-X^{+}_{\rm R}$. $\psi^{+}_{\pm}(\sigma^{\pm})$, $\psi^{-}_{0\pm}(\sigma^{\pm})$ and $\psi^{k}_{\pm}(\sigma^{\pm})$ are arbitrary periodic or antiperiodic fermionic functions of $\sigma^{\pm}$ respectively. $f(\sigma^{+})$, $g(\sigma^{-})$ are twisted bosonic functions of $\sigma^{+}$ or $\sigma^{-}$ and $\lambda_{\pm}(\sigma^{\pm})$ are twisted fermionic functions of $\sigma^{\pm}$ respectively. The twisted boundary conditions are $f(\sigma^{+}+2\pi)=e^{2\pi i\hat{\mu}}f(\sigma^{+})$, $g(\sigma^{-}-2\pi)=e^{2\pi i\hat{\mu}}g(\sigma^{-})$ and $\lambda_{+}(\sigma^{+}+2\pi)=\pm e^{2\pi i\hat{\mu}}\lambda_{+}(\sigma^{+})$, $\lambda_{-}(\sigma^{-}-2\pi)=\pm e^{2\pi i\hat{\mu}}\lambda_{-}(\sigma^{-})$ where $+$ sign denotes R sector and $-$ sign denotes NS sector. Here we define $\hat{\mu}=\mu\alpha'p^{+}$.
In (\ref{gos:-}) $X^{-}_{0\rm R}$ and $X^{-}_{0\rm L}$ are the fields which do not contain the fields $f,g,\lambda_{\pm}$ and their Hermitian conjugates however $X^{-}_{\rm 1B}$ and $X^{-}_{\rm 1F}$ contain these fields:
\begin{align}
X^{-}_{\rm 1B}&=\frac{i\mu}{2}\left[\int d\sigma^{+}
:(f^{*}\partial_{+}f-\partial_{+}f^{*}f):
-\int d\sigma^{-}:(g^{*}\partial_{-}g-\partial_{-}g^{*}g):\right]-\mu J_{\rm B}
\sigma,\label{X-1B}\\
X^{-}_{\rm 1F}&=\frac{\mu}{2}\left[\int d\sigma^{+}:\lambda^{*}_{+}\lambda_{+}:
-\int d\sigma^{-}:\lambda^{*}_{-}\lambda_{-}:\right]-\mu J_{\rm F}\sigma.
\label{X-1F}
\end{align}
Here the integrals are indefinite integrals, and we choose the constants of integration to be zero. Moreover $J_{\rm B}$ and $J_{\rm F}$ are
\begin{align}
J_{\rm B}&=\frac{i}{4\pi}\left[\int_{0}^{2\pi}d\sigma^{+}:(f^{*}\partial_{+}f-\partial_{+}f^{*}f):
+\int_{0}^{2\pi}d\sigma^{-}:(g^{*}\partial_{-}g-\partial_{-}g^{*}g):\right],
\label{JB1}\\
J_{\rm F}&=\frac{1}{4\pi}\left[\int_{0}^{2\pi}d\sigma^{+}:\lambda^{*}_{+}\lambda_{+}:
+\int_{0}^{2\pi}d\sigma^{-}:\lambda^{*}_{-}\lambda_{-}:\right]
\label{JF1},
\end{align}
where the symbol $:\ :$ denotes normal ordering which we explain later, see (\ref{normal:AB}) and (\ref{normal:lambda}).
We can expand the {\it general operator solutions} by using free modes, which are called the {\it free-mode representations}. The {\it free-mode representations} of the bosonic fields are
\begin{align}
X^{+}_{\rm L}(\sigma^{+})&=\frac{x^{+}}{2}+\frac{\alpha'}{2}p^{+}\sigma^{+}+i\sqrt{\frac{\alpha'}{2}}\sum_{n\neq 0}\frac{\tilde{\alpha}^{+}_{n}}{n}e^{-in\sigma^{+}},\ \ \ 
X^{+}_{\rm R}(\sigma^{-})=\frac{x^{+}}{2}+\frac{\alpha'}{2}p^{+}\sigma^{-}+i\sqrt{\frac{\alpha'}{2}}\sum_{n\neq 0}\frac{\alpha^{+}_{n}}{n}e^{-in\sigma^{-}},\\
X^{-}_{0\rm L}(\sigma^{+})&=\frac{x^{-}_{0}}{2}+\frac{\alpha'}{2}p^{-}_{0}\sigma^{+}+i\sqrt{\frac{\alpha'}{2}}\sum_{n\neq 0}\frac{\tilde{\alpha}^{0-}_{n}}{n}e^{-in\sigma^{+}},\ 
X^{-}_{0\rm R}(\sigma^{-})=\frac{x^{-}_{0}}{2}+\frac{\alpha'}{2}p^{-}_{0}\sigma^{-}+i\sqrt{\frac{\alpha'}{2}}\sum_{n\neq 0}\frac{\alpha^{0-}_{n}}{n}e^{-in\sigma^{-}},\\
X^{k}_{\rm L}(\sigma^{+})&=\frac{x^{k}}{2}+\frac{\alpha'}{2}p^{k}\sigma^{+}+i\sqrt{\frac{\alpha'}{2}}\sum_{n\neq 0}\frac{\tilde{\alpha}^{k}_{n}}{n}e^{-in\sigma^{+}},\ \ \ \ \ 
X^{k}_{\rm R}(\sigma^{-})=\frac{x^{k}}{2}+\frac{\alpha'}{2}p^{k}\sigma^{-}+i\sqrt{\frac{\alpha'}{2}}\sum_{n\neq 0}\frac{\alpha^{k}_{n}}{n}e^{-in\sigma^{-}},\\
f(\sigma^{+})&=\sqrt{\alpha'}\sum_{n\in\mathbb{Z}}\frac{A_{n}}{\sqrt{|n-\hat{\mu}|}}e^{-i(n-\hat{\mu})\sigma^{+}},\ \ \ \ \ \ \ \ \ \ \ \ \ \ \ \ \ \ 
g(\sigma^{-})=\sqrt{\alpha'}\sum_{n\in\mathbb{Z}}\frac{B_{n}}{\sqrt{|n+\hat{\mu}|}}e^{-i(n+\hat{\mu})\sigma^{-}},
\end{align}
where we exclude that $\hat{\mu}$ is an integer here, although we can treat it defining the solution for $\hat{\mu}$ is an integer. 
Moreover the {\it free-mode representations} of the fermionic fields are
\begin{align}
\psi^{+}_{+}(\sigma^{+})&=\sqrt{\alpha'}\sum_{r\in\mathbb{Z}+\varepsilon}\tilde{\psi}^{+}_{r}e^{-ir\sigma^{+}},\ \ \ \ \ \ \ 
\psi^{+}_{-}(\sigma^{-})=\sqrt{\alpha'}\sum_{r\in\mathbb{Z}+\varepsilon}
\psi^{+}_{r}e^{-ir\sigma^{-}},\\
\psi^{\ -}_{0+}(\sigma^{+})&=\sqrt{\alpha'}\sum_{r\in\mathbb{Z}+\varepsilon}
\tilde{\psi}^{0-}_{r}e^{-ir\sigma^{+}},\ \ \ \ \ 
\psi^{\ -}_{0-}(\sigma^{-})=\sqrt{\alpha'}\sum_{r\in\mathbb{Z}+\varepsilon}
\psi^{0-}_{r}e^{-ir\sigma^{-}},\\
\psi^{k}_{+}(\sigma^{+})&=\sqrt{\alpha'}\sum_{r\in\mathbb{Z}+\varepsilon}
\tilde{\psi}^{k}_{r}e^{-ir\sigma^{+}},\ \ \ \ \ \ \ \ 
\psi^{k}_{-}(\sigma^{-})=\sqrt{\alpha'}\sum_{r\in\mathbb{Z}+\varepsilon}
\psi^{k}_{r}e^{-ir\sigma^{-}},\\
\lambda_{+}(\sigma^{+})&=\sqrt{2\alpha'}\sum_{r\in\mathbb{Z}+\varepsilon}
\tilde{\lambda}_{r}e^{-i(r-\hat{\mu})\sigma^{+}},\ \ 
\lambda_{-}(\sigma^{-})=\sqrt{2\alpha'}\sum_{r\in\mathbb{Z}+\varepsilon}
\lambda_{r}e^{-i(r+\hat{\mu})\sigma^{-}},
\end{align}
where $\varepsilon=0$ denotes R sector and $\varepsilon=\frac{1}{2}$ denotes NS sector. Here we also exclude that $\hat{\mu}$ is a half integer because of avoidance of changing sector between R sector and NS sector.
Here the nonvanishing commutation relations between the modes of bosonic fieldes are 
\begin{align}
[x^{+},p^{-}_{0}]&=[x^{-}_{0},p^{+}]=-i,\ \ \ \ \ [\tilde{\alpha}^{+}_{m},\tilde{\alpha}^{0-}_{n}]=[\alpha^{+}_{m},\alpha^{0-}_{n}]=-m\delta_{m+n},\\
[x^{k},p^{l}]&=i\delta^{kl},\ \ \ \ \ \ \ \ \ \ \ \ \ \ \ \ \ \ \ \ [\tilde{\alpha}^{k}_{m},\tilde{\alpha}^{l}_{n}]=
[\alpha^{k}_{m},\alpha^{l}_{n}]=m\delta^{kl}\delta_{m+n},\\
[A_{m},A^{\dagger}_{n}]&={\rm sgn}(m-\hat{\mu})\delta_{m,n},\ \ \ [B_{m},B^{\dagger}_{n}]={\rm sgn}(m+\hat{\mu})\delta_{m,n}.
\end{align}
Moreover the nonvanishing anticommutation relations the modes of fermionic fields are
\begin{align}
\{\tilde{\psi}^{+}_{r},\tilde{\psi}^{0-}_{s}\}=\{\psi^{+}_{r},\psi^{0-}_{s}\}=-\delta_{r+s},\ 
\{\tilde{\psi}^{k}_{r},\tilde{\psi}^{l}_{s}\}=\{\psi^{k}_{r},\psi^{l}_{s}\}=\delta^{kl}\delta_{r+s},\ 
\{\tilde{\lambda}_{r},\tilde{\lambda}^{\dagger}_{s}\}=\{\lambda_{r},\lambda^{\dagger}_{s}\}=\delta_{r-s}.
\label{antiComRel:freemodes}
\end{align}
The commutation relations between bosonic modes and fermionic modes vanish.
The normal orderings of $A^{\dagger}_{n}A_{n},\ B^{\dagger}_{n}B_{n}$ and 
$\tilde{\lambda}^{\dagger}_{r}\tilde{\lambda}_{r},\ \lambda^{\dagger}_{r}\lambda_{r}$ are
\begin{align}
:A^{\dagger}_{n}A_{n}:&=
\begin{cases}
A^{\dagger}_{n}A_{n},\ \ (n>\hat{\mu})\\
A_{n}A^{\dagger}_{n},\ \ (n<\hat{\mu})
\end{cases},\ \ 
:B^{\dagger}_{n}B_{n}:=
\begin{cases}
B^{\dagger}_{n}B_{n},\ \ (n>-\hat{\mu})\\
B_{n}B_{n}^{\dagger},\ \ (n<-\hat{\mu})
\end{cases}\label{normal:AB},\\
:\tilde{\lambda}^{\dagger}_{r}\tilde{\lambda}_{r}:&=
\begin{cases}
\ \ \tilde{\lambda}^{\dagger}_{r}\tilde{\lambda}_{r},\ \ (r>\hat{\mu})\\
-\tilde{\lambda}_{r}\tilde{\lambda}^{\dagger}_{r},\ \ (r<\hat{\mu})
\end{cases},\ \ \ 
:\lambda^{\dagger}_{r}\lambda_{r}:=
\begin{cases}
\ \ \lambda^{\dagger}_{r}\lambda_{r},\ \ (r>-\hat{\mu})\\
-\lambda_{r}\lambda^{\dagger}_{r}.\ \ (r<-\hat{\mu})
\end{cases}\label{normal:lambda}.
\end{align}
Substituting the {\it free-mode representations} into (\ref{JB1}) and (\ref{JF1}), $J_{\rm B}$ and $J_{\rm F}$ are rewriten as 
\begin{align}
J_{\rm B}&=\alpha'\sum_{n\in\mathbb{Z}}
\Big[{\rm sgn}(n-\hat{\mu}):A^{\dagger}_{n}A_{n}:
+{\rm sgn}(n+\hat{\mu}):B^{\dagger}_{n}B_{n}:\Big],\label{JB2}\\
J_{\rm F}&=\alpha'\Big[\sum_{r\in\mathbb{Z}+\varepsilon}
:\tilde{\lambda}^{\dagger}_{r}\tilde{\lambda}_{r}:
+\sum_{r'\in\mathbb{Z}+\varepsilon}:\lambda^{\dagger}_{r'}\lambda_{r'}:\Big]\label{JF2}.
\end{align}
In the case of $J_{\rm F}$ we have to take care of the possibility that the left modes and the right modes belong the different sectors, for example the left modes belong the R sector and the right modes belong the NS sector. 
Finally we write up the super-Virasoro operator in the NS-NS pp-wave background. From now onward, we use the following modes for convinience
\begin{align}
\hat{A}_{n}=\frac{A_{n}}{\sqrt{|n-\hat{\mu}|}},\ \ 
\hat{B}_{n}=\frac{B_{n}}{\sqrt{|n+\hat{\mu}|}}.
\end{align}
In the notations we have to note that the commutation relation between $\hat{A}_{n}$, $\hat{B}_{n}$ and $x^{-}_{0}$ dose not vanish because of $[x^{-}_{0},\hat{\mu}]=-i\mu\alpha'$. The matter part of the bosonic super-Virasoro operators are
\begin{align}
\tilde{L}^{\rm M}_{n}&=\sum_{m\in\mathbb{Z}}
\left[-:\tilde{\alpha}^{+}_{n-m}\tilde{\alpha}^{0-}_{m}:
+\frac{1}{2}:\tilde{\alpha}^{k}_{n-m}\tilde{\alpha}^{k}_{m}:
+(m-n-\hat{\mu})(m-\hat{\mu}):\hat{A}^{\dagger}_{m-n}\hat{A}_{m}:\right]\nonumber\\
&\ \ \ +\sum_{r\in\mathbb{Z}+\varepsilon}\left[-2(r-\frac{n}{2})
:\tilde{\psi}^{+}_{n-r}\tilde{\psi}^{0-}_{r}:
+\frac{1}{2}(r-\frac{n}{2}):\tilde{\psi}^{k}_{n-r}\tilde{\psi}^{k}_{r}:
+(r-\frac{n}{2}-\hat{\mu}):\tilde{\lambda}^{\dagger}_{r-n}\tilde{\lambda}_{r}:\right]
\nonumber\\
&\ \ \ +\frac{\mu}{\sqrt{2\alpha'}}\tilde{\alpha}^{+}_{n}(J_{\rm B}+J_{\rm F}),
\label{Virasoro:L}\\
L^{\rm M}_{n}&=\sum_{m\in\mathbb{Z}}
\left[-:\alpha^{+}_{n-m}\alpha^{0-}_{m}:
+\frac{1}{2}:\alpha^{k}_{n-m}\alpha^{k}_{m}:
+(m-n+\hat{\mu})(m+\hat{\mu}):\hat{B}^{\dagger}_{m-n}\hat{B}_{m}\right]\nonumber\\
&\ \ \ +\sum_{r\in\mathbb{Z}+\varepsilon}\left[-(r-\frac{n}{2}):\psi^{+}_{n-r}\psi^{0-}_{r}:
+\frac{1}{2}(r-\frac{n}{2}):\psi^{k}_{n-r}\psi^{k}_{r}:
+(r-\frac{n}{2}+\hat{\mu}):\lambda^{\dagger}_{r-n}\lambda_{r}:\right]\nonumber\\
&\ \ \ -\frac{\mu}{\sqrt{2\alpha'}}\alpha^{+}_{n}(J_{\rm B}+J_{\rm F}).
\label{Virasoro:R}
\end{align}
 Moreover the matter part of the fermionic super-Virasoro operators are
\begin{align}
\tilde{G}^{\rm M}_{r}&=\sum_{n\in\mathbb{Z}}
\left[-\tilde{\psi}^{+}_{r-n}\tilde{\alpha}^{0-}_{n}
-\tilde{\psi}^{0-}_{r-n}\tilde{\alpha}^{+}_{n}
+\tilde{\psi}^{k}_{r-n}\tilde{\alpha}^{k}_{n}
+i(n-\hat{\mu})
(\tilde{\lambda}_{n+r}\hat{A}^{\dagger}_{n}
-\tilde{\lambda}^{\dagger}_{n-r}\hat{A}_{n})\right]
+\frac{\mu}{\sqrt{2\alpha'}}\tilde{\psi}^{+}_{r}(J_{\rm B}+J_{\rm F}),\label{SVirasoro:L}\\
G^{\rm M}_{r}&=\sum_{n\in\mathbb{Z}}\left[-\psi^{+}_{r-n}\alpha^{0-}_{n}-\psi^{0-}_{r-n}\alpha^{+}_{n}+\psi^{k}_{r-n}\alpha^{k}_{n}
+i(n+\hat{\mu})
(\lambda_{n+r}\hat{B}^{\dagger}_{n}-\lambda^{\dagger}_{n-r}\hat{B}_{n})\right]
-\frac{\mu}{\sqrt{2\alpha'}}\psi^{+}_{r}(J_{\rm B}+J_{\rm F}).\label{SVirasoro:R}
\end{align}
In the end of this section, we write up the ghost part of the super-Viraso operators and the BRST operator. Since the ghost modes are generally not influenced by backgronund fields, these are same modes in the flat background. They are
\begin{align}
\tilde{L}_{n}^{\rm gh}&=\sum_{m \in \mathbb{Z}}(n+m):\tilde{b}_{n-m}\tilde{c}_{m}:
+\frac{1}{2}\sum_{r\in\mathbb{Z}+\varepsilon}(3n-2r):\tilde{\gamma}_{n-r}\tilde{\beta}_{r}:,\\
{L}_{n}^{\rm gh}&=\sum_{m \in \mathbb{Z}}(n+m):{b}_{n-m}{c}_{m}:
+\frac{1}{2}\sum_{r\in\mathbb{Z}+\varepsilon}(3n-2r):{\gamma}_{n-r}{\beta}_{r}:,\\
\tilde{G}^{\rm gh}_{r}&=-2\sum_{s\in\mathbb{Z}+\varepsilon}\tilde{b}_{r-s}\tilde{\gamma}_{s}+\frac{1}{2}\sum_{s\in\mathbb{Z}+\varepsilon}(s-3r)\tilde{c}_{r-s}\tilde{\beta}_{s},\\
\tilde{G}^{\rm gh}_{r}&=-2\sum_{s\in\mathbb{Z}+\varepsilon}\tilde{b}_{r-s}\tilde{\gamma}_{s}+\frac{1}{2}\sum_{s\in\mathbb{Z}+\varepsilon}(s-3r)\tilde{c}_{r-s}\tilde{\beta}_{s}.
\end{align}
where $\tilde{b}_{n}(b_{n}),\ \tilde{c}_{n}(c_{n})$ and $\tilde{\beta}_{r}(\beta_{r}),\ \tilde{\gamma}_{r}(\gamma_{r})$ are left and right modes of the ghost fields.  
Then the BRST operator is
\begin{align}
Q_{\rm B}=&\ \ \ \sum_{m\in\mathbb{Z}}:\left[\tilde{L}^{\rm M}_{m}+\frac{1}{2}\tilde{L}^{\rm gh}_{m}-a\delta_{m,0}\right]\tilde{c}_{-m}:
+\sum_{r\in\mathbb{Z}+\varepsilon}:\left[\tilde{G}^{\rm M}_{r}+\frac{1}{2}\tilde{G}^{\rm gh}_{r}\right]\tilde{\gamma}_{-r}:\nonumber\\
&+\sum_{m\in\mathbb{Z}}:\left[{L}^{\rm M}_{m}+\frac{1}{2}{L}^{\rm gh}_{m}-a\delta_{m,0}\right]{c}_{-m}:
+\sum_{r\in\mathbb{Z}+\varepsilon}:\left[{G}^{\rm M}_{r}+\frac{1}{2}{G}^{\rm gh}_{r}\right]{\gamma}_{-r}:.\label{BRST}
\end{align} 
Here ordering constant of the super-Virasoro operator $a$ is determined by the nilpotency of the BRST operator in the NS-NS pp-wave background so that the result is 
\begin{align}
a=
\begin{cases}
0\ \ \ \ \ \ \ \ \ \ \ \ \ \ \ \ \ \ \ \ \ \ \ {\rm R\ sector}\\
|\frac{1}{2}-(\hat{\mu}-[\hat{\mu}])|\ \ \ \ {\rm NS\ sector},
\end{cases}
\end{align}
where $[x]$ is the greatest integer that is not beyond $x$, namely, the Gauss' symbol and this result is the same as our previous paper. We write up the BRST transformations of free modes in the NS-NS pp-wave background in appendix A. 


\section{Low energy string field action in the NS-NS pp-wave}
In this section, we define the low energy superstring action based on the {\it general operator solutions} and the {\it free mode representations} represented by the previous section. Next we define the low energy state, the newly defined modes and represent the low energy super-Virasoro operator by using the new modes. In the last part of this section we represent the useful formulae. 
First we define string field action as
\begin{align}
S=-\frac{1}{2}\Phi Q_{\rm B}\Phi,
\end{align}
where we do not consider the interaction terms because of simplicity.
Here the string field $\Phi$ and the BRST operator $Q_{\rm B}$ have the intricate interaction in the NS-NS pp-wave background, which is different from the case of the flat background. For closed string field, the action is defined as 
\begin{align}
S=-\frac{1}{2}\int d^{10}x\langle\Phi(x)|Q_{\rm B}b^{-}_{0}|\Phi(x)\rangle,
\end{align}
where $b^{-}_{0}$ is the ghost which vanish a half degree of freedom of closed string field. Here ghosts $b_{0}^{\pm},\ c_{0}^{\pm}$ are defined by using $\tilde{b}_{0},\ b_{0}$ and $\tilde{c}_{0},\ c_{0}$, 
\begin{align}
b^{\pm}_{0}=b_{0}\pm \tilde{b}_{0},\ c^{\pm}_{0}=c_{0}\pm\tilde{c}_{0}.
\end{align}
The non-vanishing anticommutation relation for ghosts is $\{b^{\pm}_{0},c^{\pm}_{0}\}=1$. 
The grand state $|\downarrow\downarrow\rangle$ is vanished by all the anihilation modes. Moreover these down arrows are fliped by $c^{\pm}_{0}$. Here we define $|\uparrow\uparrow\rangle=c^{+}_{0}c^{-}_{0}|\downarrow\downarrow\rangle$ and $\langle\downarrow\downarrow|\uparrow\uparrow\rangle=\langle\uparrow\uparrow|\downarrow\downarrow\rangle=1$, $\langle\downarrow\downarrow|\downarrow\downarrow\rangle=\langle\uparrow\uparrow|\uparrow\uparrow\rangle=0$. We also use the folloing notation for short: $\langle{\cal O}\rangle=\langle\downarrow\downarrow|{\cal O}|\uparrow\uparrow\rangle=\langle\uparrow\uparrow|{\cal O}|\downarrow\downarrow\rangle$, where ${\cal O}$ is an operator for something.
Here we consider the NS-NS sector of the low energy state $|\Phi(x)\rangle$ is definded by using the component fields, the low energy creation modes and the ground state:
\begin{align}
|\Phi(x)\rangle&=c^{-}_{0}\Big[
{e^{+}_{\mu\nu}}(x)
\tilde\psi^{\mu}_{-1/2}\psi^{\nu}_{-1/2}
+{\phi}(x)\tilde\beta_{-1/2}\gamma_{-1/2}
+{s}(x)\tilde\gamma_{-1/2}\beta_{-1/2}\nonumber\\ 
&\ \ \ \ \ +c^{+}_{0}\{{B_{\mu}}(x)
\tilde\beta_{-1/2}\psi^{\mu}_{-1/2}
+{E_{\mu}}(x)
\tilde\psi^{\mu}_{-1/2}\beta_{-1/2}\}\Big]
|\downarrow\downarrow\rangle,
\end{align}  
where $e^{\pm}_{\mu\nu}(x)=h_{\mu\nu}(x)\pm b_{\mu\nu}(x)$. Here $h_{\mu\nu}(x)$ denotes the gravitational field, $b_{\mu\nu}(x)$ denotes the NS-NS antisymmetric tensor field and $\phi(x),\ s(x)$ denote the scalar fields and $B_{\mu},\ E_{\mu}$ denote auxiliary fields. Moreover $\tilde{\psi}^{\mu}_{-1/2},\ \psi^{\mu}_{-1/2}$ are low energy modes which we explain later (\ref{mode:zL})-(\ref{mode:-R}), and $\tilde{\beta}_{-1/2},\ \beta_{-1/2}$ and $\tilde{\gamma}_{-1/2},\ \gamma_{-1/2}$ are ghosts whose commutation relations are $[\tilde{\gamma}_{\pm 1/2},\tilde{\beta}_{\mp 1/2}]=1,\ [\gamma_{\pm 1/2},\beta_{\mp 1/2}]=1$. Usually the state is expanded by using the infinite number of component fields, however we particularly notice the low energy fields, which is the most characteristic point. Moreover we have to note that this construction is based on the paper by Berkovits \cite{Berkovits}, however in our construction, the low energy state is defined in the NS-NS pp-wave background, so that the modes $\tilde{\psi}^{\mu}_{-1/2},\ \psi^{\mu}_{-1/2}$ are influenced by the background. The state $\langle\Phi(x)|$ is defined as Hermitian conjugate of $|\Phi(x)\rangle$:
\begin{align}
\langle\Phi(x)|&=\langle\downarrow\downarrow|\Big[\psi^{\alpha}_{1/2}
\tilde{\psi}^{\beta}_{1/2}e^{-}_{\alpha\beta}(x)
-\gamma_{1/2}\tilde{\beta}_{1/2}\phi(x)
-\beta_{1/2}\tilde{\gamma}_{1/2}s(x)\nonumber\\
&\ \ \ \ \ +\{\psi^{\alpha}_{1/2}\tilde{\beta}_{1/2}B^{\alpha}(x)
+\beta_{1/2}\tilde{\psi}^{\alpha}_{1/2}E^{\alpha}(x)\}c^{+}_{0}\Big]c^{-}_{0}.
\end{align}
Here $e^{+}_{\mu\nu}$ terns into $e^{-}_{\mu\nu}$ because of the property of antisymmetric tensor field $b_{\mu\nu}$ and the minus sign in front of the scalar fields comes from anti-Hermiticity of $\tilde{\beta}_{r},\ \beta_{r}$ whose Hermitian conjugates are $\tilde{\beta}_{r}^{\dagger}=-\tilde{\beta}_{-r},\ \beta_{r}^{\dagger}=-\beta_{-r}$ (e.g. see \cite{IMNU}).    
The BRST operater (\ref{BRST}) becomes the following form in this low energy case:
\begin{align}
Q_{\rm B}&=\left[\tilde{L}_{0}^{\rm M}+L^{\rm M}_{0}-1+2\hat{\mu}
-\frac{1}{2}(\tilde{\gamma}_{-1/2}\tilde{\beta}_{1/2}
-\tilde{\beta}_{-1/2}\tilde{\gamma}_{1/2}
+\gamma_{-1/2}\beta_{1/2}
-\beta_{-1/2}\gamma_{1/2})\right]c_{0}^{+}\nonumber\\
&+\tilde{G}^{\rm M}_{-1/2}\tilde{\gamma}_{1/2}
+\tilde{G}^{\rm M}_{1/2}\tilde{\gamma}_{-1/2}
+{G}^{\rm M}_{-1/2}{\gamma}_{1/2}
+{G}^{\rm M}_{1/2}{\gamma}_{-1/2}
-(\tilde{\gamma}_{-1/2}\tilde{\gamma}_{1/2}
+\gamma_{-1/2}\gamma_{1/2})b^{+}_{0}
+\cdots,
\end{align}
where the last term of $\cdots$ means high energy parts. From now on we restrict the eigen value of $\hat{\mu}$ to a real number in $0<\hat{\mu}<\frac{1}{2}$ because of the simplicity.
In the modes $\tilde{\psi}^{\mu}_{\pm 1/2}, \psi^{\mu}_{\pm 1/2}$ which are contained in the low energy states, the components of $\mu=+,k$ are same definition as free mode representation, however the components of $\mu=z,z^{*},-$ are defined as
\begin{align}
\tilde\psi^{z}_{\pm 1/2}&=\sqrt{2}\tilde{\lambda}_{\pm 1/2}
-i\mu{z}\tilde\psi^{+}_{\pm1/2},\label{mode:zL}\\
\psi^{z}_{\pm 1/2}&=\sqrt{2}\lambda_{\pm 1/2}
+i\mu{z}\psi^{+}_{\pm1/2},\label{mode:zR}\\
\tilde\psi^{-}_{\pm 1/2}&=\tilde\psi^{0-}_{\pm 1/2}
-\frac{i\mu}{\sqrt{2}}[{z}\tilde\lambda^{\dagger}_{\mp 1/2}
-{z^{*}}\tilde\lambda_{\pm 1/2}],\label{mode:-L}\\
\psi^{-}_{\pm 1/2}&=\psi^{0-}_{\pm 1/2}
+\frac{i\mu}{\sqrt{2}}[{z}\lambda^{\dagger}_{\mp 1/2}
-{z^{*}}\lambda_{\pm 1/2}],\label{mode:-R}
\end{align}
where $z^{*}$ component is obtained by taking Hermitian conjugate of $z$ component. The details of the definition of these modes are explained in Appendix B, especially (\ref{psi^z_s}) and (\ref{psi^-_s}).
In these equation, the coordinate dependence comes from
\begin{align}
\hat{A}_{0}+\hat{B}_{0}=\frac{1}{\sqrt{\alpha'}}z,\ \ \hat{A}^{\dagger}_{0}+\hat{B}^{\dagger}_{0}=\frac{1}{\sqrt{\alpha'}}z^{*},
\end{align}
whose relations also explained in (\ref{COM:zz*}).
We have to note that these modes depend on the coordinates of $z,\ z^{*}$ linearly. In the case of more general backgrounds, we expect that modes depend on more general functions of coordinates.
The nonvanishing anticommutation relations of these modes are culculated by using anticommutation relations of the free modes (\ref{antiComRel:freemodes}), 
\begin{align}
\{\tilde{\psi}^{\mu}_{\pm 1/2},\tilde{\psi}^{\nu}_{\mp 1/2}\}=\{\psi^{\mu}_{\pm 1/2},\psi^{\nu}_{\mp 1/2}\}=g^{\mu\nu},
\label{antiComRel:modes}
\end{align}
which become contravariant metric of NS-NS pp-wave background (\ref{pp-wave:metric}).
Using the new modes (\ref{mode:zL})-(\ref{mode:-R}) the bosonic super-Virasoro operator (\ref{Virasoro:L}), (\ref{Virasoro:R}) and the fermionic super-Virasoro operator (\ref{SVirasoro:L}), (\ref{SVirasoro:R}) become the following form in the low energy case:
\begin{align}
\tilde L^{\rm M}_{0}+L^{\rm M}_{0}=-\frac{\alpha'}{2}
(g^{\rho\sigma}
\partial_{\rho}\partial_{\sigma}+2i\mu\partial_{-})
+\frac{1}{2}g_{\rho\sigma}
(:\tilde{\psi}^{\rho}_{-1/2}\tilde{\psi}^{\sigma}_{1/2}:
+:{\psi}^{\rho}_{-1/2}{\psi}^{\sigma}_{1/2}:)
-i\mu K_{\rm F}\partial_{-}
+\cdots,\label{Virasoro:LowEne}
\end{align}
\begin{align}
\tilde G^{\rm M}_{\pm 1/2}=\sqrt{\frac{\alpha'}{2}}
(-i\tilde\psi^{\mu}_{\pm 1/2}\partial_{\mu}
+\frac{\mu}{\alpha'}
\tilde\psi^{+}_{\pm 1/2}J_{\rm F})
+\cdots,\ \ 
G^{\rm M}_{\pm 1/2}=\sqrt{\frac{\alpha'}{2}}
(-i\psi^{\mu}_{\pm 1/2}\partial_{\mu}-\frac{\mu}{\alpha'}
\psi^{+}_{\pm 1/2}J_{\rm F})
+\cdots,\label{SVirasoro:LowEne}
\end{align}
where $\cdots$ means high energy parts and $J_{\rm B}$ is canceled out in this representation. Here the new operator $K_{\rm F}$ is defined as 
\begin{align}
K_{\rm F}=\alpha'\sum_{r=\pm 1/2}
\left[:\tilde\lambda^{\dagger}_{r}\tilde\lambda_{r}:
-:\lambda^{\dagger}_{r}\lambda_{r}:\right].\label{KF}
\end{align}
Although the operator $J_{\rm F}$ in eq. (\ref{JF2}) is defined by summation over all the harf integer, the new operator $K_{\rm F}$ is defined by summation over only $\pm 1/2$. However looking at the low energy level, the difference between $J_{\rm F}$ and $K_{\rm F}$ is the sign of the right moving modes so that we can calculate the commutation relation for $K_{\rm F}$ using the commutaion relation for $J_{\rm F}$
\begin{align}
[K_{\rm F},\tilde{\psi}^{\mu}_{\pm 1/2}]&=+[J_{\rm F},\tilde{\psi}^{\mu}_{\pm 1/2}],\label{ComRel:KF1}\\
[K_{\rm F},\psi^{\mu}_{\pm 1/2}]&=-[J_{\rm F},\psi^{\mu}_{\pm 1/2}]\label{ComRel:KF2}.
\end{align}
Moreover we have to note the definition of differential operator $\partial_{\mu}$, namely zero mode of bosonic fields:
\begin{align}
\tilde{\alpha}^{+}_{0}&=\alpha^{+}_{0}=i\sqrt{\frac{\alpha'}{2}}\partial_{-},\ 
\tilde{\alpha}^{0-}_{0}=\alpha^{0-}_{0}=i\sqrt{\frac{\alpha'}{2}}\partial_{+},\ 
\tilde{\alpha}^{k}_{0}=\alpha^{k}_{0}=-i\sqrt{\frac{\alpha'}{2}}\partial_{k},\\ 
\hat{A}_{0}&=\frac{1}{2\sqrt{\alpha'}}(z-2\alpha'\hat{\mu}^{-1}\partial_{z*}),\ \ 
\hat{B}_{0}=\frac{1}{2\sqrt{\alpha'}}(z+2\alpha'\hat{\mu}^{-1}\partial_{z*}),\\
\hat{A}_{0}^{\dagger}&=\frac{1}{2\sqrt{\alpha'}}(z^{*}+2\alpha'\hat{\mu}^{-1}\partial_{z}),\ \
\hat{B}_{0}^{\dagger}=\frac{1}{2\sqrt{\alpha'}}(z^{*}-2\alpha'\hat{\mu}^{-1}\partial_{z}),
\end{align}
where the operators $\hat{A}_{0},\ \hat{B}_{0}$ and their Hermitian conjugates are harmonic oscillator which contain coordinates explained in (\ref{A0B0}). 

In the end of this section, we explain the outline for the following section whose perpose is calculation of the action by using component fields specifically. The super-Virasoro operator in the BRST operator contain the differential operator $\partial_{\mu}$ and the special operators namely $J_{\rm F}$ and $K_{\rm F}$. According to the definition of the action, these operators operate on the state which contain $\tilde{\psi}^{\mu}_{\pm 1/2},\ \psi^{\mu}_{\pm 1/2}$.
Since these modes have linear dependence on the coordinates $z,\ z^{*}$, we have to note the differential operator $\partial_{z},\ \partial_{z*}$ so that $\partial_{\mu}$ operates on not only the fields but also the modes.
Moreover we have to calculate that $J_{\rm F}$ and $K_{\rm F}$ operate on the modes.
After this culculation, finally we take expectation value by using (\ref{antiComRel:modes}), namely they become $\langle\tilde{\psi}^{\mu}_{1/2}\tilde{\psi}^{\nu}_{-1/2}\rangle=\langle\psi^{\mu}_{1/2}\psi^{\nu}_{-1/2}\rangle=g^{\mu\nu}$.
The series of these calculations leads us to the action of component fields.
In these calculations, we had better write up some useful formulae.
First after differentiating the modes (\ref{mode:zL})-(\ref{mode:-R}), we represent the modes without $\tilde{\lambda}_{\pm 1/2},\ \lambda_{\pm 1/2}$ and their Hermitian conjugates by using (\ref{mode:zL}), (\ref{mode:zR}) again. The results are 
\begin{align}
\partial_{\alpha}\tilde{\psi}^{\mu}_{\pm 1/2}=
&-\frac{i\mu}{2}
(\tilde{\psi}^{z*}_{\pm 1/2}-i\mu z^{*}\tilde{\psi}^{+}_{\pm 1/2})
\delta^{z}_{\alpha}\delta^{\mu}_{-}
-i\mu\tilde{\psi}^{+}_{\pm 1/2}\delta^{z}_{\alpha}\delta^{\mu}_{z}\nonumber\\
&+\frac{i\mu}{2}
(\tilde{\psi}^{z}_{\pm 1/2}+i\mu z\tilde{\psi}^{+}_{\pm 1/2})
\delta^{z*}_{\alpha}\delta^{\mu}_{-}
+i\mu\tilde{\psi}^{+}_{\pm 1/2}\delta^{z*}_{\alpha}\delta^{\mu}_{z*},\label{der:psi:L}\\
\partial_{\alpha}\psi^{\mu}_{\pm 1/2}=&
\frac{i\mu}{2}
(\psi^{z*}_{\pm 1/2}+i\mu z^{*}\psi^{+}_{\pm 1/2})
\delta^{z}_{\alpha}\delta^{\mu}_{-}
+i\mu\psi^{+}_{\pm 1/2}\delta^{z}_{\alpha}\delta^{\mu}_{z}\nonumber\\
&-\frac{i\mu}{2}
(\psi^{z}_{\pm 1/2}-i\mu z\psi^{+}_{\pm 1/2})
\delta^{z*}_{\alpha}\delta^{\mu}_{-}
-i\mu\psi^{+}_{\pm 1/2}\delta^{z*}_{\alpha}\delta^{\mu}_{z*}.\label{der:psi:R}
\end{align}
Therefore the following expectation values are calculated as 
\begin{align}
\langle\tilde{\psi}^{\mu}_{1/2}\partial_{\rho}\tilde{\psi}^{\nu}_{-1/2}\rangle
=&-\frac{i\mu}{2}(g^{\mu z*}-i\mu z^{*}g^{\mu+})\delta^{z}_{\rho}\delta^{\nu}_{-}-i\mu g^{\mu+}\delta^{z}_{\rho}\delta^{\nu}_{z}\nonumber\\
&+\frac{i\mu}{2}(g^{\mu z}+i\mu zg^{\mu+})\delta^{z*}_{\rho}\delta^{\nu}_{-}
+i\mu g^{\mu+}\delta^{z*}_{\rho}\delta^{\nu}_{z*},\label{Ex:psi:L}\\
\langle\psi^{\mu}_{1/2}\partial_{\rho}\psi^{\nu}_{-1/2}\rangle
=&\frac{i\mu}{2}(g^{\mu z*}+i\mu z^{*}g^{\mu+})\delta^{z}_{\rho}\delta^{\nu}_{-}+i\mu g^{\mu+}\delta^{z}_{\rho}\delta^{\nu}_{z}\nonumber\\
&-\frac{i\mu}{2}(g^{\mu z}-i\mu zg^{\mu+})\delta^{z*}_{\rho}\delta^{\nu}_{-}
-i\mu g^{\mu+}\delta^{z*}_{\rho}\delta^{\nu}_{z*}.\label{Ex:psi:R}
\end{align}
Since the expectation value of modes represents the inverse metric, these differential operator sandwiched between modes roughly represent like the Christoffel symbol from the standpoint of superstiring field theory.
In the next place, we represent the fomulae of the commutation relation between $J_{\rm F}$ and modes, according to the relations (\ref{ComRel:KF1}) and (\ref{ComRel:KF2}), we need not represent the fomulae on $K_{\rm F}$:
\begin{align}
[J_{\rm F},\tilde{\psi}^{\mu}_{\pm 1/2}]&=\alpha'\left\{
-\frac{i\mu}{2}(z\tilde{\psi}^{z*}_{\pm 1/2}+z^{*}\tilde{\psi}^{z}_{\pm 1/2})\delta^{\mu}_{-}
-(\tilde{\psi}^{z}_{\pm 1/2}+i\mu z\tilde{\psi}^{+}_{\pm 1/2})\delta^{\mu}_{z}
+(\tilde{\psi}^{z*}_{\pm 1/2}-i\mu z^{*}\tilde{\psi}^{+}_{\pm 1/2})\delta^{\mu}_{z*}\right\},\label{ComRel:JF:L}\\
[J_{\rm F},\tilde{\psi}^{\mu}_{\pm 1/2}]&=\alpha'\left\{
\frac{i\mu}{2}(z\psi^{z*}_{\pm 1/2}+z^{*}\psi^{z}_{\pm 1/2})\delta^{\mu}_{-}
-(\psi^{z}_{\pm 1/2}-i\mu z\psi^{+}_{\pm 1/2})\delta^{\mu}_{z}
+(\psi^{z*}_{\pm 1/2}+i\mu z^{*}\psi^{+}_{\pm 1/2})\delta^{\mu}_{z*}\right\}.\label{ComRel:JF:R}
\end{align} 
Moreover the expectation value become as follows:
\begin{align}
\langle\tilde{\psi}^{\beta}_{1/2}[J_{\rm F},\tilde{\psi}^{\mu}_{-1/2}]\rangle
&=\alpha'\left\{-\frac{i\mu}{2}(zg^{\beta z*}+z^{*}g^{\beta z})\delta^{\mu}_{-}
-(g^{\beta z}+i\mu zg^{\beta+})\delta^{\mu}_{z}
+(g^{\beta z*}-i\mu z^{*}g^{\beta+})\delta^{\mu}_{z*}\right\},\label{ExComRel:JF:L}\\
\langle\psi^{\alpha}_{1/2}[J_{\rm F},\psi^{\nu}_{-1/2}]\rangle
&=\alpha'\left\{\frac{i\mu}{2}(zg^{\alpha z*}+z^{*}g^{\alpha z})\delta^{\nu}_{-}-(g^{\alpha z}-i\mu zg^{\alpha+})\delta^{\nu}_{z}+(g^{\alpha z*}+i\mu z^{*}g^{\alpha+})\delta^{\nu}_{z*}\right\}.\label{ExComRel:JF:R}
\end{align}
Using these fomulae we can easily calculate the action of component fields.


\section{Component representation of superstring field action}
In this section we specifically calculate the components of superstiring field action in the NS-NS pp-wave background. On the calcualtion, we use the modes based on the {\it general operator solutions} and some useful formulae of these expectation values constructed in the previous section. Calculating the expectation values we can obtain the action for graviton, antisymmetric tensor field and dilaton influenced by the NS-NS pp-wave background. In the next section we will confirm whether this action correspond to the second-order perturbation of supergravity action from the NS-NS pp-wave background.
From the relation between creation operator and annihilation operator for ghosts, we have already known the remaining modes so that we had better calculate the expectation value for the ghost modes first. Using the (anti)commutation relation for ghosts, we can eliminate ghost modes from the action. 
Let us start from the superstring field action 
\begin{align}
S=-\frac{1}{2}\int d^{10}x\Big[
&-\langle\psi^{\alpha}_{1/2}
\tilde\psi^{\beta}_{1/2} 
e^{-}_{\alpha\beta}
(\tilde L_{0}^{\rm M}+L_{0}^{\rm M}-1+2\hat{\mu})
e^{+}_{\mu\nu}\tilde\psi^{\mu}_{-1/2}\psi^{\nu}_{-1/2}\rangle\nonumber\\
&+\langle\phi(\tilde L_{0}^{\rm M}+L_{0}^{\rm M}-1+2\hat{\mu})s\rangle
+\langle s(\tilde L_{0}^{\rm M}+L_{0}^{\rm M}-1+2\hat{\mu})\phi\rangle
+2\langle s\phi\rangle\nonumber\\
&+\langle\psi^{\alpha}_{1/2}\tilde{\psi}^{\beta}_{1/2}e^{-}_{\alpha\beta}
\tilde{G}^{\rm M}_{-1/2}B_{\mu}\psi^{\mu}_{-1/2}\rangle
+\langle\psi^{\alpha}_{1/2}\tilde{\psi}^{\beta}_{1/2}e^{-}_{\alpha\beta}
G^{\rm M}_{-1/2}E_{\mu}\tilde{\psi}^{\mu}_{-1/2}\rangle\nonumber\\
&+\langle sG^{\rm M}_{1/2}B_{\mu}\psi^{\mu}_{-1/2}\rangle
+\langle\phi\tilde{G}^{\rm M}_{1/2}E_{\mu}\tilde{\psi}^{\mu}_{-1/2}\rangle
\nonumber\\
&-\langle\psi^{\alpha}_{1/2}B_{\alpha}\tilde{G}^{\rm M}_{1/2}e^{+}_{\mu\nu}
\tilde{\psi}^{\mu}_{-1/2}\psi^{\nu}_{-1/2}\rangle
-\langle\tilde{\psi}^{\alpha}_{1/2}E_{\alpha}G^{\rm M}_{1/2}e^{+}_{\mu\nu}
\tilde{\psi}^{\mu}_{-1/2}\psi^{\nu}_{-1/2}\rangle\nonumber\\
&-\langle\tilde{\psi}^{\alpha}_{1/2}E_{\alpha}\tilde{G}^{\rm M}_{-1/2}\phi\rangle-\langle\psi^{\alpha}_{1/2}B_{\alpha}G^{\rm M}_{-1/2}s\rangle\nonumber\\
&+\langle\tilde{\psi}^{\alpha}_{1/2}B_{\alpha}B_{\mu}
\tilde{\psi}^{\mu}_{-1/2}\rangle
+\langle\psi^{\alpha}_{1/2}E_{\alpha}E_{\mu}
\psi^{\mu}_{-1/2}\rangle
\Big],
\end{align}
which we define in the previous section by using component fields without ghost modes.
Here the operator ordering constant $\hat{\mu}=\mu\alpha'p^{+}=i\alpha'\mu\partial_{-}$ vanish from the action because we can make total derivative from this operator. In the case of the two scalar fields we can make total derivative easily. $\phi\hat{\mu}s+s\hat{\mu}\phi=i\alpha'\mu\partial_{-}(\phi s)$. In the case of the tensor fields $e^{\pm}_{\mu\nu}$, because of the world sheet fermions $\psi^{\mu}_{\pm 1/2}$ not depending on $x^{-}$, we can calculate the expectation value easily. Here we use that the expectation value of modes become contravariant metric, $\langle\tilde{\psi}^{\mu}_{1/2}\tilde{\psi}^{\nu}_{-1/2}\rangle=\langle\psi^{\mu}_{1/2}\psi^{\nu}_{-1/2}\rangle=g^{\mu\nu}$.
The calculation is 
\begin{align}
\langle\psi^{\alpha}_{1/2}
\tilde\psi^{\beta}_{1/2} 
e^{-}_{\alpha\beta}
\hat{\mu}
e^{+}_{\mu\nu}\tilde\psi^{\mu}_{-1/2}\psi^{\nu}_{-1/2}\rangle
&=i\alpha'\mu g^{\alpha\nu}g^{\beta\mu}e^{-}_{\alpha\beta}\partial_{-}
e^{+}_{\mu\nu}\nonumber\\
&=i\alpha'\mu\partial_{-}(h^{\mu\nu}h_{\mu\nu}+b^{\mu\nu}b_{\mu\nu}).
\end{align}
Above calculation the cross terms $h^{\mu\nu}\partial_{-}b_{\mu\nu}$ and $b^{\mu\nu}\partial_{-}h_{\mu\nu}$ vanish automatically because of the product of the symmetry and the antisymmetry of spacetime indices.
Moreover using the following identities,
\begin{align}
\int d^{10}x\langle\psi^{\alpha}_{1/2}\tilde{\psi}^{\beta}_{1/2}e^{-}_{\alpha\beta}\tilde{G}^{\rm M}_{-1/2}B_{\mu}\psi^{\mu}_{-1/2}\rangle
&=-\int d^{10}x\langle\psi^{\alpha}_{1/2}B_{\alpha}
\tilde{G}^{\rm M}_{1/2}e^{+}_{\mu\nu}\tilde{\psi}^{\mu}_{-1/2}\psi^{\nu}_{-1/2}
\rangle,\\
\int d^{10}x\langle\psi^{\alpha}_{1/2}\tilde{\psi}^{\beta}_{1/2}e^{-}_{\alpha\beta}G^{\rm M}_{-1/2}E_{\mu}\tilde{\psi}^{\mu}_{-1/2}\rangle
&=-\int d^{10}x\langle\tilde{\psi}^{\alpha}_{1/2}E_{\alpha}
G^{\rm M}_{1/2}e^{+}_{\mu\nu}\tilde{\psi}^{\mu}_{-1/2}\psi^{\nu}_{-1/2}
\rangle,\\
\int d^{10}x\langle sG^{\rm M}_{1/2}B_{\mu}\psi^{\mu}_{-1/2}\rangle
&=-\int d^{10}x\langle\psi^{\alpha}_{1/2}B_{\alpha}G^{\rm M}_{-1/2}s\rangle,\\
\int d^{10}x\langle\phi\tilde{G}^{\rm M}_{1/2}E_{\mu}\tilde{\psi}^{\mu}_{-1/2}
\rangle
&=-\int d^{10}x\langle\tilde{\psi}^{\alpha}_{1/2}E_{\alpha}\tilde{G}^{\rm M}_{-1/2}\phi\rangle,
\end{align}
the action becomes simpler form. Here we can prove these identities from the Hermiticity of the action and taking Hermitian conjugate of left hand side. The minus sign of right hand side comes from the ghosts $c_{0}^{\pm}$ hidden in expectation value. We can also prove them by direct calculation of the following subsection. Using the identities the action becomes
\begin{align}
S=-\frac{1}{2}&\int d^{10}x\Big[-\langle\psi^{\alpha}_{1/2}
\tilde\psi^{\beta}_{1/2} 
e^{-}_{\alpha\beta}
(\tilde L_{0}^{\rm M}+L_{0}^{\rm M}-1)e^{+}_{\mu\nu}
\tilde\psi^{\mu}_{-1/2}\psi^{\nu}_{-1/2}\rangle\nonumber\\
&+\langle\phi(\tilde L_{0}^{\rm M}+L_{0}^{\rm M})s\rangle
+\langle s(\tilde L_{0}^{\rm M}+L_{0}^{\rm M})\phi\rangle
\nonumber\\
&-2{B_{\alpha}}\{{\langle\psi^{\alpha}_{1/2}
\tilde{G}^{\rm M}_{1/2}e^{+}_{\mu\nu}
\tilde{\psi}^{\mu}_{-1/2}\psi^{\nu}_{-1/2}\rangle
+\langle\psi^{\alpha}_{1/2}G^{\rm M}_{-1/2}s\rangle}\}
\nonumber\\
&-2{E_{\alpha}}\{\langle\tilde{\psi}^{\alpha}_{1/2}
G^{\rm M}_{1/2}e^{+}_{\mu\nu}
\tilde{\psi}^{\mu}_{-1/2}\psi^{\nu}_{-1/2}\rangle
+\langle\tilde{\psi}^{\alpha}_{1/2}\tilde{G}^{\rm M}_{-1/2}
\phi\rangle\}\nonumber\\
&+{B_{\mu}B^{\mu}}+{E_{\mu}E^{\mu}}\Big].
\end{align}
From the action we can derive the equation of motion for $B^{\mu},\ E^{\mu}$ and we substitute the solutions of the equation of motion for the action. Then removing the auxiliary fields $B^{\mu},\ E^{\mu}$, the action becomes
\begin{align}
S=-\frac{1}{2}&\int d^{10}x\Big[-\langle\psi^{\alpha}_{1/2}
\tilde\psi^{\beta}_{1/2} 
e^{-}_{\alpha\beta}
(\tilde L_{0}^{\rm M}+L_{0}^{\rm M}
-1)e^{+}_{\mu\nu}
\tilde\psi^{\mu}_{-1/2}\psi^{\nu}_{-1/2}\rangle\nonumber\\
&+\langle\phi(\tilde L_{0}^{\rm M}+L_{0}^{\rm M})s\rangle
+\langle s(\tilde L_{0}^{\rm M}+L_{0}^{\rm M})\phi\rangle
-(\bar{B}_{\mu}\bar{B}^{\mu}+
\bar{E}_{\mu}\bar{E}^{\mu})\Big],
\end{align}
where we newly define $\bar{B}^{\mu},\ \bar{E}^{\mu}$ as 
\begin{align}
\bar{B}^{\alpha}&=\langle\psi^{\alpha}_{1/2}
\tilde{G}^{\rm M}_{1/2}e^{+}_{\mu\nu}
\tilde{\psi}^{\mu}_{-1/2}\psi^{\nu}_{-1/2}\rangle
+\langle\psi^{\alpha}_{1/2}G^{\rm M}_{-1/2}s\rangle,\\
\bar{E}^{\alpha}&=\langle\tilde{\psi}^{\alpha}_{1/2}
G^{\rm M}_{1/2}e^{+}_{\mu\nu}
\tilde{\psi}^{\mu}_{-1/2}\psi^{\nu}_{-1/2}\rangle
+\langle\tilde{\psi}^{\alpha}_{1/2}\tilde{G}^{\rm M}_{-1/2}
\phi\rangle.
\end{align}
Here we have to note that $\bar{B}^{\mu}$ and $\bar{E}^{\mu}$ are no longer auxiliary fields. 
Moreover using gauge symmetry of $s(x)$ we can remove $s(x)$ always where we discuss the gauge symmetry in section 6, especially (\ref{eta}) and (\ref{GT2}). Finally the action becomes the simplest form:
\begin{align}
S=-\frac{1}{2}&\int d^{10}x\left[-\langle\psi^{\alpha}_{1/2}
\tilde\psi^{\beta}_{1/2} 
e^{-}_{\alpha\beta}
(\tilde L_{0}^{\rm M}+L_{0}^{\rm M}
-1)e^{+}_{\mu\nu}
\tilde\psi^{\mu}_{-1/2}\psi^{\nu}_{-1/2}\rangle
-(\bar{B}_{\mu}\bar{B}^{\mu}+
\bar{E}_{\mu}\bar{E}^{\mu})\right]\label{simplest action}.
\end{align}
In this case the kinetic term of the dilaton field comes out from the term of 
$\bar{E}_{\mu}\bar{E}^{\mu}$ as we understand it later.
\subsection{Calculation of the bosonic super-Virasoro operator part}
In this subsection we calculate the action of the bosonic super-Virasoro operator part. 
We name this term $S_{1}$:
\begin{align}
S_{1}=-\frac{1}{2}&\int d^{10}x\left[-\langle\psi^{\alpha}_{1/2}
\tilde\psi^{\beta}_{1/2} 
e^{-}_{\alpha\beta}
(\tilde L_{0}^{\rm M}+L_{0}^{\rm M}
-1)e^{+}_{\mu\nu}
\tilde\psi^{\mu}_{-1/2}\psi^{\nu}_{-1/2}\rangle\right].
\end{align}
As we discuss in the previous section, we can use the bosonic super-Virasoro operator (\ref{Virasoro:LowEne}) in the case of treating the low energy case.
In this operator the term $2\mu i\partial_{-}$ and $\frac{1}{2}g_{\rho\sigma}(:\tilde{\psi}^{\rho}_{-1/2}\tilde{\psi}^{\sigma}_{1/2}:+:\psi^{\rho}_{-1/2}\psi^{\sigma}_{1/2}:)$ vanish because the former becomes total derivative and the latter and ordering constant $(-1)$ cancel out each other so that practically we calculate the first term and the forth term in the bosonic super-Virasoro operator so that $S_{1}$ becomes
\begin{align}
S_{1}=-\frac{1}{2}\int d^{10}x\left[-\langle\psi^{\alpha}_{1/2}
\tilde\psi^{\beta}_{1/2} 
e^{-}_{\alpha\beta}
(-\frac{\alpha'}{2}g^{\rho\sigma}\partial_{\rho}\partial_{\sigma}
-i\mu K_{\rm F}\partial_{-})
e^{+}_{\mu\nu}
\tilde\psi^{\mu}_{-1/2}\psi^{\nu}_{-1/2}\rangle\right].
\end{align}
In this calculation the most important point is that $\tilde{\psi}^{\mu}_{-1/2}$ and $\psi^{\mu}_{-1/2}$ are the linear functions of the coordinates of $z$ or $z^{*}$ so that we have to pay attention to the differential operators $\partial_{z}, \partial_{z*}$. Therefore we have to multiply $\tilde{\psi}^{\mu}_{-1/2}$ and $\psi^{\mu}_{-1/2}$ by differential operator $\partial_{\mu}$ at only one time. The first term becomes
\begin{align}
\langle\psi^{\alpha}_{1/2}
&\tilde\psi^{\beta}_{1/2} 
e^{-}_{\alpha\beta}
(-\frac{\alpha'}{2}g^{\rho\sigma}\partial_{\rho}\partial_{\sigma})
e^{+}_{\mu\nu}
\tilde\psi^{\mu}_{-1/2}\psi^{\nu}_{-1/2}\rangle\nonumber\\
&=-\frac{\alpha'}{2}g^{\rho\sigma}e^{-}_{\alpha\beta}
\langle\psi^{\alpha}_{1/2}
\tilde\psi^{\beta}_{1/2}(\partial_{\rho}\partial_{\sigma}
e^{+}_{\mu\nu}
\tilde\psi^{\mu}_{-1/2}\psi^{\nu}_{-1/2}
+2\partial_{\rho}e^{+}_{\mu\nu}\partial_{\sigma}\tilde{\psi}^{\mu}_{-1/2}\psi^{\nu}_{-1/2}\nonumber\\
&\ \ \ +2\partial_{\rho}e^{+}_{\mu\nu}\tilde{\psi}^{\mu}_{-1/2}
\partial_{\sigma}\psi^{\nu}
+2e^{+}_{\mu\nu}\partial_{\rho}\tilde{\psi}^{\mu}_{-1/2}\partial_{\sigma}\psi^{\nu}_{-1/2})\rangle.
\end{align}
Then we calculate the expectation value of modes without differential operator 
on ahead as
\begin{align}
\langle\psi^{\alpha}_{1/2}
&\tilde\psi^{\beta}_{1/2} 
e^{-}_{\alpha\beta}
(-\frac{\alpha'}{2}g^{\rho\sigma}\partial_{\rho}\partial_{\sigma})
e^{+}_{\mu\nu}
\tilde\psi^{\mu}_{-1/2}\psi^{\nu}_{-1/2}\rangle\nonumber\\
&=-\frac{\alpha'}{2}\bigg\{
g^{\rho\sigma}g^{\alpha\nu}g^{\beta\mu}
e^{-}_{\alpha\beta}\partial_{\rho}\partial_{\sigma}e^{+}_{\mu\nu}
+2g^{\rho\sigma}e^{-}_{\alpha\beta}\partial_{\rho}e^{+}_{\mu\nu}
\left(g^{\alpha\nu}\langle\tilde{\psi}^{\beta}_{1/2}\partial_{\sigma}\tilde{\psi}^{\mu}_{-1/2}\rangle+g^{\beta\mu}\langle\psi^{\alpha}_{1/2}\partial_{\sigma}\psi^{\nu}_{-1/2}\rangle\right)\nonumber\\
&\ \ \ \ \ \ \ \ \ \ \ \ +2g^{\rho\sigma}e^{-}_{\alpha\beta}e^{+}_{\mu\nu}
\langle\tilde{\psi}^{\beta}_{1/2}\partial_{\sigma}\tilde{\psi}^{\mu}_{-1/2}
\rangle\langle\psi^{\alpha}_{1/2}\partial_{\sigma}\psi^{\nu}_{-1/2}\rangle
\bigg\}.\label{S1-1}
\end{align}
In the next place we calculate the term of $K_{\rm F}$. The modes $\tilde{\psi}^{\mu}_{-1/2}$ and $\psi^{\mu}_{-1/2}$ do not depend on $x^{-}$ so that we only multiply $\partial_{-}$ by $e^{\pm}_{\mu\nu}$, finally we take expectation value on ahead. The caluculation is
\begin{align}
\langle\psi^{\alpha}_{1/2}&\tilde{\psi}^{\beta}_{1/2}e^{-}_{\alpha\beta}
(-i\mu K_{\rm F}\partial_{-})
e^{+}_{\mu\nu}\tilde{\psi}^{\mu}_{-1/2}\psi^{\nu}_{-1/2}\rangle\nonumber\\
&=-i\mu e^{-}_{\alpha\beta}\partial_{-}e^{+}_{\mu\nu}
\langle\psi^{\alpha}_{1/2}\tilde{\psi}^{\beta}_{1/2}K_{\rm F}\tilde{\psi}^{\mu}
_{-1/2}\psi^{\nu}_{-1/2}\rangle\nonumber\\
&=-i\mu e^{-}_{\alpha\beta}\partial_{-}e^{+}_{\mu\nu}
\left\{\langle\psi^{\alpha}_{1/2}\tilde{\psi}^{\beta}_{1/2}
[K_{\rm F},\tilde{\psi}^{\mu}_{-1/2}]\psi^{\nu}_{-1/2}\rangle
+\langle\psi^{\alpha}_{1/2}\tilde{\psi}^{\beta}_{1/2}\tilde{\psi}^{\mu}_{-1/2}
[K_{\rm F},\psi^{\nu}_{-1/2}]\rangle\right\}\nonumber\\
&=-i\mu e^{-}_{\alpha\beta}\partial_{-}e^{+}_{\mu\nu}
\left\{g^{\alpha\nu}\langle\tilde{\psi}^{\beta}_{1/2}
[J_{\rm F},\tilde{\psi}^{\mu}_{-1/2}]\rangle
-g^{\beta\mu}\langle\psi^{\alpha}[J_{\rm F},\psi^{\nu}_{-1/2}]\rangle\right\},
\label{S1-2}
\end{align}
where we use the relation (\ref{ComRel:KF1}) and (\ref{ComRel:KF2}) in the last line of this calculation. Moreover expectation values $\langle\tilde{\psi}^{\mu}_{1/2}\partial_{\rho}\tilde{\psi}^{\nu}_{-1/2}\rangle$, $\langle{\psi}^{\mu}_{1/2}\partial_{\rho}{\psi}^{\nu}_{-1/2}\rangle$ and $\langle\tilde{\psi}^{\mu}_{1/2}[J_{\rm F},\tilde{\psi}^{\nu}_{-1/2}]\rangle$, $\langle{\psi}^{\mu}_{1/2}[J_{\rm F},{\psi}^{\nu}_{-1/2}]\rangle$ in (\ref{S1-1}) and (\ref{S1-2}) whose formulae are constructed in (\ref{der:psi:L})-(\ref{ExComRel:JF:R}). 
Although we feel a little complicated, they are not so difficult because of the formulae almost constructed by Kronecker's delta and the inverse metric of NS-NS pp-wave.
Thus using these expectation values, we can calculate the addition $g^{\alpha\nu}\langle\tilde{\psi}^{\beta}_{1/2}\partial_{\sigma}\tilde{\psi}^{\mu}_{-1/2}\rangle+g^{\beta\mu}\langle\psi^{\alpha}_{1/2}\partial_{\sigma}
\psi^{\nu}_{-1/2}\rangle$ and the product $g^{\rho\sigma}\langle\tilde{\psi}^{\alpha}_{1/2}\partial_{\rho}\tilde{\psi}^{\mu}_{-1/2}\rangle\langle\psi^{\beta}_{1/2}\partial_{\sigma}\psi^{\nu}_{-1/2}\rangle$. The addition becomes 
\begin{align}
g^{\alpha\nu}&\langle\tilde{\psi}^{\beta}_{1/2}\partial_{\sigma}\tilde{\psi}^{\mu}_{-1/2}\rangle+g^{\beta\mu}\langle\psi^{\alpha}_{1/2}\partial_{\sigma}
\psi^{\nu}_{-1/2}\rangle\nonumber\\
&=\ \ \ g^{\alpha\nu}\left[\frac{i\mu}{2}(g^{\beta z}+i\mu zg^{\beta+})
\delta^{z*}_{\sigma}\delta^{\mu}_{-}-\frac{i\mu}{2}(g^{\beta z*}-i\mu z^{*}
g^{\beta+})\delta^{z}_{\sigma}\delta^{\mu}_{-}+i\mu g^{\beta+}
(\delta^{z*}_{\sigma}\delta^{\mu}_{z*}-\delta^{z}_{\sigma}\delta^{\mu}_{z})\right]\nonumber\\
&\ \ \ +g^{\beta\mu}\left[\frac{i\mu}{2}(g^{\alpha z*}+i\mu z^{*}g^{\alpha+})
\delta^{z}_{\sigma}\delta^{\nu}_{-}-\frac{i\mu}{2}(g^{\alpha z}-i\mu zg^{\alpha+})\delta^{z*}_{\sigma}\delta^{\nu}_{-}+i\mu g^{\alpha+}
(\delta^{z}_{\sigma}\delta^{\nu}_{z}-\delta^{z*}_{\sigma}\delta^{\nu}_{z*})\right].
\end{align}
We can simplify this, if we replace the spacetime indeces with common indeces using the (anti)symmetry of spacetime indeces of tensor fields $e^{\pm}_{\mu\nu}$, then
\begin{align}
\langle\tilde{\psi}^{\mu}_{1/2}\partial_{\rho}\tilde{\psi}^{\nu}_{-1/2}\rangle+\langle{\psi}^{\mu}_{1/2}\partial_{\rho}{\psi}^{\nu}_{-1/2}\rangle&=-\mu^{2}g^{\mu+}\delta^{\nu}_{-}(z\delta^{z*}_{\rho}+z^{*}\delta^{z}_{\rho}),\\
\langle\tilde{\psi}^{\mu}_{1/2}\partial_{\rho}\tilde{\psi}^{\nu}_{-1/2}\rangle-\langle{\psi}^{\mu}_{1/2}\partial_{\rho}{\psi}^{\nu}_{-1/2}\rangle&=i\mu(g^{\mu z}\delta^{z*}_{\rho}-g^{\mu z*}\delta^{z}_{\rho})\delta^{\nu}_{-}+2i\mu g^{\mu+}
(\delta^{z*}_{\rho}\delta^{\nu}_{z*}-\delta^{z}_{\rho}\delta^{\nu}_{z}).
\end{align}
The product becomes 
\begin{align}
g^{\rho\sigma}
&\langle\tilde{\psi}^{\alpha}_{1/2}\partial_{\rho}\tilde{\psi}^{\mu}_{-1/2}
\rangle
\langle\psi^{\beta}_{1/2}\partial_{\sigma}\psi^{\nu}_{-1/2}\rangle\nonumber\\
&=\frac{\mu^{2}}{2}\bigg[
\big\{g^{z(\alpha}g^{\beta)z*}+i\mu zg^{+[\alpha}g^{\beta]z*}
-i\mu z^{*}g^{+[\alpha}g^{\beta]z}-2\mu^{2}z^{*}zg^{+\alpha}g^{+\beta}\big\}
\delta^{\mu}_{-}\delta^{\nu}_{-}\nonumber\\
&\ \ \ \ \ \ \ \ \ \ +2\big\{(g^{\alpha z}\delta^{\nu}_{z}+g^{\alpha z*}\delta^{\nu}_{z*})\big\}
g^{\beta+}\delta^{\nu}_{-}
+(g^{\beta z}\delta^{\mu}_{z}+g^{\beta z*}\delta^{\mu}_{z*})
g^{\alpha+}\delta^{\nu}_{-}
+i\mu g^{+\alpha}g^{+\beta}(z\delta^{[\mu}_{-}\delta^{\nu]}_{z}-z^{*}
\delta^{[\mu}_{-}\delta^{\nu]}_{z*})\big\}\nonumber\\
&\ \ \ \ \ \ \ \ \ \ +4g^{+\alpha}g^{+\beta}\delta^{(\mu}_{z}\delta^{\nu)}_{z*}
\bigg].\label{product:<><>}
\end{align}
Here we define the following (anti)symmetric symbols,  
\begin{align}
g^{\mu(\alpha}g^{\beta)\nu}&=g^{\mu\alpha}g^{\beta\nu}+g^{\mu\beta}g^{\alpha\nu},\ \ 
g^{\mu[\alpha}g^{\beta]\nu}=g^{\mu\alpha}g^{\beta\nu}-g^{\mu\beta}g^{\alpha\nu},\\
\delta^{(\mu}_{\alpha}\delta^{\nu)}_{\beta}
&=\delta^{\mu}_{\alpha}\delta^{\nu}_{\beta}
+\delta^{\nu}_{\alpha}\delta^{\mu}_{\beta},\ \ \ \ \ \ \ \ \ \ \ 
\delta^{[\mu}_{\alpha}\delta^{\nu]}_{\beta}
=\delta^{\mu}_{\alpha}\delta^{\nu}_{\beta}
-\delta^{\nu}_{\alpha}\delta^{\mu}_{\beta},
\end{align}
because of the avoidance of the complexity and using the symbols we are also easy to see the property of symmetry of spacetime indeces.
From now onward we divide $e^{\pm}_{\mu\nu}$ into $h_{\mu\nu}$ and $b_{\mu\nu}$ for using symmetry and antisymmetry of spacetime indeces. We represent $e^{-}_{\alpha\beta}e^{+}_{\mu\nu}$, $e^{-}_{\alpha\beta}\partial_{\rho}e^{+}_{\mu\nu}$ and $e^{-}_{\alpha\beta}\partial_{\rho}\partial_{\sigma}e^{+}_{\mu\nu}$ using $h_{\mu\nu}$ and $b_{\mu\nu}$,
\begin{align}
e^{-}_{\alpha\beta}e^{+}_{\mu\nu}&=h_{\alpha\beta}h_{\mu\nu}+h_{\alpha\beta}b_{\mu\nu}-b_{\alpha\beta}h_{\mu\nu}-b_{\alpha\beta}b_{\mu\nu},\\
e^{-}_{\alpha\beta}\partial_{\rho}e^{+}_{\mu\nu}&=h_{\alpha\beta}\partial_{\rho}h_{\mu\nu}+h_{\alpha\beta}\partial_{\rho}b_{\mu\nu}-b_{\alpha\beta}\partial_{\rho}h_{\mu\nu}-b_{\alpha\beta}\partial_{\rho}b_{\mu\nu},\\
e^{-}_{\alpha\beta}\partial_{\rho}\partial_{\sigma}e^{+}_{\mu\nu}&=h_{\alpha\beta}\partial_{\rho}\partial_{\sigma}h_{\mu\nu}+h_{\alpha\beta}\partial_{\rho}\partial_{\sigma}b_{\mu\nu}-b_{\alpha\beta}\partial_{\rho}\partial_{\sigma}h_{\mu\nu}-b_{\alpha\beta}\partial_{\rho}\partial_{\sigma}b_{\mu\nu}.
\end{align}
We evaluate the term of $e^{-}_{\alpha\beta}e^{+}_{\mu\nu}g^{\rho\sigma}
\langle\tilde{\psi}^{\alpha}_{1/2}\partial_{\rho}\tilde{\psi}^{\mu}_{-1/2}\rangle
\langle\psi^{\beta}_{1/2}\partial_{\sigma}\psi^{\nu}_{-1/2}\rangle$ dividing $e_{\mu\nu}^{\pm}$ into $h_{\mu\nu}$ and $b_{\mu\nu}$.
It is important that because of (anti)symmetry of spacetime indeces in the tensor fields, different terms appear in (\ref{product:<><>}): 
\begin{align}
&h_{\alpha\beta}h_{\mu\nu}g^{\rho\sigma}\langle\tilde{\psi}^{\alpha}_{1/2}\partial_{\rho}\tilde{\psi}^{\mu}_{-1/2}\rangle
\langle\psi^{\beta}_{1/2}\partial_{\sigma}\psi^{\nu}_{-1/2}\rangle\nonumber\\
&\ \ \ \ \ \ =-\frac{\mu^{2}}{2}h_{\alpha\beta}h_{\mu\nu}\Big[2\{g^{\alpha z}g^{\beta z*}-\mu^{2}z^{*}z g^{\alpha+}g^{\beta+}\}\delta^{\mu}_{-}\delta^{\nu}_{-}+2\{(g^{\alpha z*}\delta^{\nu}_{z*}+g^{\alpha z}\delta^{\nu}_{z})g^{\beta+}\delta^{\mu}_{-}\}
\nonumber\\
&\ \ \ \ \ \ \ \ \ \ \ \ \ \ \ \ \ \ \ \ \ \ \ \ \ \ \ +2\{(g^{\beta z}\delta^{\mu}_{z}+g^{\beta z*}\delta^{\mu}_{z*})g^{\alpha+}\delta^{\nu}_{-}\}
+8g^{\alpha+}g^{\beta+}\delta^{\mu}_{z}\delta^{\nu}_{z*}\Big],\label{hh<><>}\\
&b_{\alpha\beta}h_{\mu\nu}g^{\rho\sigma}\langle\tilde{\psi}^{\alpha}_{1/2}\partial_{\rho}\tilde{\psi}^{\mu}_{-1/2}\rangle
\langle\psi^{\beta}_{1/2}\partial_{\sigma}\psi^{\nu}_{-1/2}\rangle
=-\frac{\mu^{2}}{2}b_{\alpha\beta}h_{\mu\nu}\cdot 2i\mu g^{\alpha+}(zg^{\beta z*}-z^{*}g^{\beta z})\delta^{\mu}_{-}\delta^{\nu}_{-},\\
&h_{\alpha\beta}b_{\mu\nu}g^{\rho\sigma}\langle\tilde{\psi}^{\alpha}_{1/2}\partial_{\rho}\tilde{\psi}^{\mu}_{-1/2}\rangle
\langle\psi^{\beta}_{1/2}\partial_{\sigma}\psi^{\nu}_{-1/2}\rangle
=-\frac{\mu^{2}}{2}h_{\alpha\beta}b_{\mu\nu}\cdot 4i\mu g^{\alpha+}g^{\beta+}\delta^{\mu}_{-}
(z\delta^{\nu}_{z}-z^{*}\delta^{\nu}_{z*}),\\
&b_{\alpha\beta}b_{\mu\nu}g^{\rho\sigma}\langle\tilde{\psi}^{\alpha}_{1/2}\partial_{\rho}\tilde{\psi}^{\mu}_{-1/2}\rangle
\langle\psi^{\beta}_{1/2}\partial_{\sigma}\psi^{\nu}_{-1/2}\rangle
=-\frac{\mu^{2}}{2}b_{\alpha\beta}b_{\mu\nu}\cdot 2\left[(g^{\alpha z*}\delta^{\nu}_{z*}+g^{\alpha z}\delta^{\nu}_{z})g^{\beta+}\delta^{\mu}_{-}
+(g^{\beta z}\delta^{\mu}_{z}+g^{\beta z*}\delta^{\mu}_{z*})g^{\alpha+}\delta^{\nu}_{-}\right]\label{bb<><>}.
\end{align}
Moreover we evaluate $\langle\psi^{\alpha}_{1/2}\tilde{\psi}^{\beta}_{1/2}e^{-}_{\alpha\beta}(-i\mu K_{\rm F}\partial_{-})
e^{+}_{\mu\nu}\tilde{\psi}^{\mu}_{-1/2}\psi^{\nu}_{-1/2}\rangle$ dividing $e^{\pm}_{\mu\nu}$ into $h_{\mu\nu}$ and $b_{\mu\nu}$:
\begin{align}
\langle\psi^{\alpha}_{1/2}\psi^{\beta}_{1/2}h_{\alpha\beta}(-i\mu K_{\rm F}\partial_{-})
h_{\mu\nu}\tilde{\psi}^{\mu}_{-1/2}\psi^{\nu}_{-1/2}\rangle
&=4\alpha'\mu^{2}g^{\beta\nu}h_{-\beta}(z\partial_{-}h_{z\nu}+z^{*}\partial_{-}h_{z*\nu})\nonumber\\
&\ \ \ \ -\partial_{-}\left\{2\alpha'\mu^{2}g^{\beta\nu}
(zh_{z\beta}h_{-\nu}+z^{*}h_{z*\beta}h_{-\nu})\right\},\label{hKh}\\
\langle\psi^{\alpha}_{1/2}\tilde{\psi}^{\beta}_{1/2}(-b_{\alpha\beta})(-i\mu K_{\rm F}\partial_{-})
h_{\mu\nu}\tilde{\psi}^{\mu}_{-1/2}\psi^{\nu}_{-1/2}\rangle
&=4\alpha'i\mu g^{\beta\nu}(b_{z*\beta}\partial_{-}h_{z\nu}-b_{z\beta}\partial_{-}h_{z*\beta}),\\
\langle\psi^{\alpha}_{1/2}\tilde{\psi}^{\beta}_{1/2}h_{\alpha\beta}(-i\mu K_{\rm F}\partial_{-})
b_{\mu\nu}\tilde{\psi}^{\mu}_{-1/2}\psi^{\nu}_{-1/2}\rangle
&=4\alpha'i\mu g^{\beta\nu}(h_{z*\beta}\partial_{-}b_{z\nu}-
h_{z\beta}\partial_{-}b_{z*\nu}),\\
\langle\psi^{\alpha}_{1/2}\tilde{\psi}^{\beta}_{1/2}(-b_{\alpha\beta})(-i\mu K_{\rm F}\partial_{-})
b_{\mu\nu}\tilde{\psi}^{\mu}_{-1/2}\psi^{\nu}_{-1/2}\rangle
&=4\alpha'\mu^{2}g^{\beta\nu}b_{-\beta}(z\partial_{-}b_{z\nu}+z^{*}\partial_{-}b_{z*\nu})\nonumber\\
&\ \ \ \ -\partial_{-}\left\{2\alpha'\mu^{2}g^{\beta\nu}(zh_{z\beta}b_{-\nu}
+z^{*}h_{z*\beta}b_{-\nu})\right\}.\label{bKb}
\end{align}
Using the formulae (\ref{hh<><>})-(\ref{bb<><>}) and (\ref{hKh})-(\ref{bKb}), we can evaluate the components of $S_{1}$:
\begin{align}
&\langle\psi^{\alpha}_{1/2}\tilde{\psi}^{\beta}_{1/2}h_{\alpha\beta}(\tilde{L}^{\rm M}_{0}+L^{\rm M}_{0}-1)h_{\mu\nu}\tilde{\psi}^{\mu}_{-1/2}\psi^{\nu}_{-1/2}\rangle\nonumber\\
&=-\frac{\alpha'}{2}\Big[h^{\mu\nu}g^{\rho\sigma}\partial_{\rho}\partial_{\sigma}h_{\mu\nu}-4\mu^{2}h^{+\mu}
(z\partial_{z}h_{\mu-}+z^{*}\partial_{z*}h_{\mu-})-8\mu^{2}g^{\beta\nu}h_{-\beta}(z\partial_{-}h_{z\nu}+z^{*}\partial_{-}h_{z*\nu})\nonumber\\
&\ \ \ \ \ \ \ \ \ \ \ \ \ -4\mu^{2}(h^{+z}h_{-z}+h^{+z*}h_{-z*})-16\mu^{2}h^{++}h_{zz*}+2\mu^{4}z^{*}zh^{++}h_{--}
\Big]\nonumber\\
&\ \ \ \ \ \ \ \ \ \ \ \ \ +\partial_{-}\{\alpha'\mu^{2}g^{\beta\nu}(zh^{z*}_{\beta}h^{+}_{\nu}+z^{*}h^{z}_{\beta}h^{+}_{\nu})\},\label{S1:hh}\\
&\langle\psi^{\alpha}_{1/2}\tilde{\psi}^{\beta}_{1/2}(-b_{\alpha\beta})(\tilde{L}^{\rm M}_{0}+L^{\rm M}_{0}-1)h_{\mu\nu}\tilde{\psi}^{\mu}_{-1/2}\psi^{\nu}_{-1/2}\rangle\nonumber\\
&=-\frac{\alpha'}{2}\Big[-g^{\rho\sigma}g^{\alpha\nu}g^{\beta\mu}b_{\alpha\beta}\partial_{\rho}\partial_{\sigma}h_{\mu\nu}+8i\mu g^{\beta\nu}(b_{z*\beta}\partial_{z}h_{-\nu}-b_{z\beta}\partial_{z*}h_{-\nu}-b_{-\beta}\partial_{z}h_{z*\nu}
+b_{-\beta}\partial_{z*}h_{z\nu})\nonumber\\
&\ \ \ \ \ \ \ \ \ \ \ \ \ +4i\mu^{3}h_{--}(zb_{-z}-z^{*}b_{-z*})-8i\mu g^{\beta\nu}(b_{z*\beta}\partial_{-}h_{z\nu}-b_{z\beta}\partial_{-}h_{z*\nu})
\Big]\label{S1:bh},\\
&\langle\psi^{\alpha}_{1/2}\tilde{\psi}^{\beta}_{1/2}h_{\alpha\beta}(\tilde{L}^{\rm M}_{0}+L^{\rm M}_{0}-1)b_{\mu\nu}\tilde{\psi}^{\mu}_{-1/2}\psi^{\nu}_{-1/2}\rangle\nonumber\\
&=-\frac{\alpha'}{2}\Big[g^{\rho\sigma}g^{\alpha\nu}g^{\beta\mu}h_{\alpha\beta}\partial_{\rho}\partial_{\sigma}b_{\mu\nu}+8i\mu g^{\beta\nu}(h_{z*\beta}\partial_{z}b_{-\nu}-h_{z\beta}\partial_{z*}b_{-\nu}-h_{-\beta}\partial_{z}b_{z*\nu}
+h_{-\beta}\partial_{z*}b_{z\nu})\nonumber\\
&\ \ \ \ \ \ \ \ \ \ \ \ \ -4i\mu^{3}h_{--}(zb_{-z}-z^{*}b_{-z*})-8i\mu g^{\beta\nu}(h_{z*\beta}\partial_{-}b_{z\nu}-h_{z\beta}\partial_{-}b_{z*\nu})
\Big]\label{S1:hb},\\
&\langle\psi^{\alpha}_{1/2}\tilde{\psi}^{\beta}_{1/2}(-b_{\alpha\beta})(\tilde{L}^{\rm M}_{0}+L^{\rm M}_{0}-1)b_{\mu\nu}\tilde{\psi}^{\mu}_{-1/2}\psi^{\nu}_{-1/2}\rangle\nonumber\\
&=-\frac{\alpha'}{2}\Big[-g^{\rho\sigma}g^{\alpha\nu}g^{\beta\mu}b_{\alpha\beta}\partial_{\rho}\partial_{\sigma}b_{\mu\nu}+4\mu^{2}g^{\beta\nu}b_{-\beta}(z\partial_{z}b_{-\nu}+z^{*}\partial_{z*}b_{-\nu})\nonumber\\
&\ \ \ \ \ \ \ \ \ \ \ \ \ -16\mu^{2}b_{-z}b_{-z*}-8\mu^{2}g^{\beta\nu}b_{-\beta}(z\partial_{-}b_{z\nu}+z^{*}\partial_{-}b_{z*\nu})
\Big]-\partial_{-}\{2\alpha'\mu^{2}g^{\beta\nu}(zb_{z\beta}b_{-\nu}+z^{*}b_{z*\beta}b_{-\nu})\},
\label{S1:bb}
\end{align}
where the terms of $\partial_{-}$ times fields come from the terms of $K_{\rm F}\partial_{-}$ and the total derivative is not important because we can integrate out it. Moreover we had better perform partial integral on (\ref{S1:bh}) for comparizon with supergravity. The result is
\begin{align}
&\langle\psi^{\alpha}_{1/2}\tilde{\psi}^{\beta}_{1/2}(-b_{\alpha\beta})(\tilde{L}^{\rm M}_{0}+L^{\rm M}_{0}-1)h_{\mu\nu}\tilde{\psi}^{\mu}_{-1/2}\psi^{\nu}_{-1/2}\rangle\nonumber\\
&=-\frac{\alpha'}{2}\Big[-g^{\rho\sigma}g^{\alpha\nu}g^{\beta\mu}b_{\alpha\beta}\partial_{\rho}\partial_{\sigma}h_{\mu\nu}\nonumber\\
&\ \ \ \ \ \ \ \ \ \ \ \ \ +8i\mu g^{\beta\nu}\big\{h_{-\nu}(\partial_{z*}b_{z\beta}-\partial_{z}b_{z*\beta})+h_{z\nu}(\partial_{-}b_{z*\beta}-\partial_{z*}b_{-\beta})+h_{z*\nu}(\partial_{z}b_{-\beta}-\partial_{-}b_{z\beta})\big\}\nonumber\\
&\ \ \ \ \ \ \ \ \ \ \ \ \ -4i\mu^{3}h_{--}(zb_{-z}-z^{*}b_{-z*})
\Big]+({\rm total\ derivative})\label{S1:bh2}.
\end{align}
Here we use the following identity, 
\begin{align}
\partial_{\mu}g^{\rho\sigma}=\mu^{2}(z\delta_{\mu}^{z*}+z^{*}\delta_{\mu}^{z})\delta^{\rho}_{-}\delta^{\sigma}_{-},
\label{der:metric}
\end{align}
for performing the partial integral.
Finally we can write up $S_{1}$ as the following form: 
\begin{align}
S_{1}=-\frac{\alpha'}{4}\int d^{10}x\Big\{
&h^{\mu\nu}g^{\rho\sigma}\partial_{\rho}\partial_{\sigma}h_{\mu\nu}-4\mu^{2}h^{+\mu}(z\partial_{z}h_{\mu-}+z^{*}\partial_{z*}h_{\mu-})-8\mu^{2}g^{\beta\nu}h_{-\beta}(z\partial_{-}h_{z\nu}+z^{*}\partial_{-}h_{z*\nu})\nonumber\\
&-4\mu^{2}(h^{+z}h_{-z}+h^{+z*}h_{-z*})-16\mu^{2}h^{++}h_{zz*}+2\mu^{4}z^{*}zh^{++}h_{--}\nonumber\\
&-g^{\rho\sigma}g^{\alpha\nu}g^{\beta\mu}b_{\alpha\beta}\partial_{\rho}\partial_{\sigma}b_{\mu\nu}
+4\mu^{2}g^{\beta\nu}b_{-\beta}(z\partial_{z}b_{-\nu}+z^{*}\partial_{z*}b_{-\nu})\nonumber\\
&-8\mu^{2}g^{\beta\nu}b_{-\beta}(z\partial_{-}b_{z\nu}+z^{*}\partial_{-}b_{z*\nu})-16\mu^{2}b_{-z}b_{-z*}\nonumber\\
&+16i\mu g^{\beta\nu}\left[h_{-\beta}(\partial_{z*}b_{z\nu}-\partial_{z}b_{z*\nu})
+h_{z\beta}(\partial_{-}b_{z*\nu}-\partial_{z*}b_{-\nu})
+h_{z*\beta}(\partial_{z}b_{-\nu}-\partial_{-}b_{z\nu})\right]\nonumber\\
&-8i\mu^{3}h_{--}(zb_{-z}-z^{*}b_{-z*})
\Big\},\label{S1:final}
\end{align}
where we can prove $\int d^{10}x g^{\rho\sigma}g^{\alpha\nu}g^{\beta\mu}(h_{\alpha\beta}\partial_{\rho}\partial_{\sigma}b_{\mu\nu}-b_{\alpha\beta}\partial_{\rho}\partial_{\sigma}h_{\mu\nu})=0$, using partial integral. We have to note that we can represent $S_{1}$ in various ways, namely using partial integral, we can represent different forms.
\subsection{Calculation of the fermionic super-Virasoro operator part}
In this subsection we calculate the action of $\bar{B}_{\mu}\bar{B}^{\mu}$ and $\bar{E}_{\mu}\bar{E}^{\mu}$ which contain the fermionic super-Virasoro operators $\tilde{G}^{\rm M}_{-1/2}$ and  $G^{\rm M}_{-1/2}$. We name this term $S_{2}$:
\begin{align}
S_{2}=-\frac{1}{2}\int d^{10}x\left[-(\bar{B}_{\mu}\bar{B}^{\mu}+\bar{E}_{\mu}\bar{E}^{\mu})\right].
\end{align}
First we calculate the square of $\bar{B}^{\alpha}$ and $\bar{E}^{\alpha}$. In the case of fixing the gauge $s(x)=0$, the square of $\bar{B}^{\alpha}$ is easy but the square of $\bar{E}^{\alpha}$ is a little complicated because of exsistence of $\phi$, then
\begin{align}
\bar{B}_{\alpha}\bar{B}^{\alpha}
&=g_{\alpha\beta}
\langle\psi^{\alpha}_{1/2}\tilde{G}_{1/2}^{\rm M}e^{+}_{\mu\nu}
\tilde{\psi}^{\mu}_{-1/2}\psi^{\nu}_{-1/2}\rangle
\langle\psi^{\beta}_{1/2}\tilde{G}_{1/2}^{\rm M}e^{+}_{\rho\sigma}
\tilde{\psi}^{\rho}_{-1/2}\psi^{\sigma}_{-1/2}\rangle,\label{BB0}\\
\bar{E}_{\alpha}\bar{E}^{\alpha}&=g_{\alpha\beta}
\langle\tilde{\psi}^{\alpha}_{1/2}G_{1/2}^{\rm M}e^{+}_{\mu\nu}
\tilde{\psi}^{\mu}_{-1/2}\psi^{\nu}_{-1/2}\rangle
\langle\tilde{\psi}^{\beta}_{1/2}G_{1/2}^{\rm M}e^{+}_{\rho\sigma}
\tilde{\psi}^{\rho}_{-1/2}\psi^{\sigma}_{-1/2}\rangle\nonumber\\
&+2g_{\alpha\beta}
\langle\tilde{\psi}^{\alpha}_{1/2}G_{1/2}^{\rm M}e^{+}_{\mu\nu}
\tilde{\psi}^{\mu}_{-1/2}\psi^{\nu}_{-1/2}\rangle
\langle\tilde{\psi}^{\beta}_{1/2}\tilde{G}^{\rm M}_{-1/2}\phi\rangle
+g_{\alpha\beta}\langle\tilde{\psi}^{\alpha}_{1/2}
\tilde{G}^{\rm M}_{-1/2}\phi\rangle
\langle\tilde{\psi}^{\beta}_{1/2}\tilde{G}^{\rm M}_{-1/2}\phi\rangle\label{EE0}.
\end{align}
As we discuss in the previous section, we can use (\ref{SVirasoro:LowEne}) as fermionic super-Virasoro operators in the low energy case.  Seeing the terms of $\bar{B}_{\alpha}\bar{B}^{\alpha}$ and $\bar{E}_{\alpha}\bar{E}^{\alpha}$ in (\ref{BB0}) and (\ref{EE0}), first of all we have to calculate $\langle\psi^{\alpha}_{1/2}\tilde{G}^{\rm M}_{1/2}e^{+}_{\mu\nu}
\tilde{\psi}^{\mu}_{-1/2}\psi^{\nu}_{-1/2}\rangle$ and $\langle\tilde{\psi}^{\alpha}_{1/2}\tilde{G}^{\rm M}_{-1/2}\phi\rangle$.
 Here we can calculate $\langle\tilde{\psi}^{\alpha}_{1/2}{G}^{\rm M}_{1/2}e^{+}_{\mu\nu}\tilde{\psi}^{\mu}_{-1/2}\psi^{\nu}_{-1/2}\rangle$ in $\bar{E}_{\alpha}\bar{E}^{\alpha}$ by using the same calculation as  $\langle\psi^{\alpha}_{1/2}\tilde{G}^{\rm M}_{1/2}e^{+}_{\mu\nu}
\tilde{\psi}^{\mu}_{-1/2}\psi^{\nu}_{-1/2}\rangle$ in $\bar{B}_{\alpha}\bar{B}^{\alpha}$ so that we show the calculation of 
 $\langle\psi^{\alpha}_{1/2}\tilde{G}^{\rm M}_{1/2}e^{+}_{\mu\nu}
\tilde{\psi}^{\mu}_{-1/2}\psi^{\nu}_{-1/2}\rangle$ representatively. 
 First we can calculate $\langle\tilde{\psi}^{\alpha}_{1/2}\tilde{G}^{\rm M}_{-1/2}\phi\rangle$ easily because $\phi$ dose not contain world sheet fermion so that $J_{\rm F}$ in $G^{\rm M}_{-1/2}$ vanish by the ground state. Moreover the derivative in $\tilde{G}^{\rm M}_{-1/2}$ operates scalar field $\phi$ and expectation values of world sheet fermions become the background contravariant metric $\langle\tilde{\psi}^{\mu}_{1/2}\tilde{\psi}^{\nu}_{-1/2}\rangle=\langle\psi^{\mu}_{1/2}\psi^{\nu}_{-1/2}\rangle=g^{\mu\nu}$. The calculation is 
\begin{align}
\langle\tilde{\psi}^{\alpha}_{1/2}\tilde{G}^{\rm M}_{-1/2}\phi\rangle
&=\sqrt{\frac{\alpha'}{2}}\langle\tilde{\psi}^{\alpha}_{1/2}
\left[-i\tilde{\psi}^{\lambda}_{-1/2}\partial_{\lambda}+\frac{\mu}{\alpha'}
\psi^{+}_{-1/2}J_{\rm F}\right]\phi\rangle\nonumber\\
&=-i\sqrt{\frac{\alpha'}{2}}\langle\tilde{\psi}^{\alpha}_{1/2}\tilde{\psi}^{\lambda}_{-1/2}
\partial_{\lambda}\phi\rangle
=-i\sqrt{\frac{\alpha'}{2}}g^{\alpha\lambda}\partial_{\lambda}\phi.
\end{align}
In the next place we calculate $\langle\psi^{\alpha}_{1/2}\tilde{G}^{\rm M}_{1/2}e^{+}_{\mu\nu}\tilde{\psi}^{\mu}_{-1/2}\psi^{\nu}_{-1/2}\rangle$. In this calculation the derivative operator $\partial_{\mu}$ in the fermionic super-Virasoro operators multiplies not only $e^{+}_{\mu\nu}$ but also $\tilde{\psi}^{\mu}_{-1/2}$ and $\psi^{\nu}_{-1/2}$ because of the dependence on the coordinates $z, z^{*}$.  Moreover we have to calculate the product between  $J_{\rm F}$ and  
$\tilde{\psi}^{\mu}_{-1/2}$, ($\psi^{\nu}_{-1/2}$). We can calculate them using the commutation relations. Finally we calculate the expectation values of fermions which become background contravariant metric. The calculation is@
\begin{align}
\langle\psi^{\alpha}_{1/2}\tilde{G}^{\rm M}_{1/2}e^{+}_{\mu\nu}\tilde{\psi}^{\mu}_{-1/2}\psi^{\nu}_{-1/2}\rangle
&=\sqrt{\frac{\alpha'}{2}}\langle\psi^{\alpha}_{1/2}
\left[
-i\tilde{\psi}^{\lambda}_{1/2}\partial_{\lambda}+\frac{\mu}{\alpha'}\tilde{\psi}^{+}_{1/2}J_{\rm F}
\right]
e^{+}_{\mu\nu}\tilde{\psi}^{\mu}_{-1/2}\psi^{\nu}_{-1/2}\rangle\nonumber\\
&=\sqrt{\frac{\alpha'}{2}}
\left\{
-i\langle\psi^{\alpha}_{1/2}\tilde{\psi}^{\lambda}_{1/2}\partial_{\lambda}
(e^{+}_{\mu\nu}\tilde{\psi}^{\mu}_{-1/2}\psi^{\nu}_{-1/2})\rangle
+\frac{\mu}{\alpha'}e^{+}_{\mu\nu}\langle\psi^{\alpha}_{1/2}
\tilde{\psi}^{+}_{1/2}[J_{\rm F},\tilde{\psi}^{\mu}_{-1/2}\psi^{\nu}_{-1/2}]
\rangle
\right\}\nonumber\\
&=\sqrt{\frac{\alpha'}{2}}
\Big\{
-ig^{\alpha\nu}g^{\lambda\mu}\partial_{\lambda}e^{+}_{\mu\nu}
-ie^{+}_{\mu\nu}(g^{\alpha\nu}\langle\tilde{\psi}^{\lambda}_{1/2}\partial_{\lambda}\tilde{\psi}^{\mu}_{-1/2}\rangle+g^{\lambda\mu}
\langle\psi^{\alpha}_{1/2}\partial_{\lambda}\psi^{\nu}_{-1/2}\rangle\nonumber\\
&\ \ \ \ \ \ \ \ \ \ \ \ \ +i\frac{\mu}{\alpha'}g^{\alpha\nu}\langle\tilde{\psi}^{+}_{1/2}[J_{\rm F},\tilde{\psi}^{\mu}_{-1/2}]\rangle
+i\frac{\mu}{\alpha'}g^{+\mu}\langle\psi^{\alpha}_{1/2}[J_{\rm F},\psi^{\nu}_{-1/2}]\rangle)
\Big\}.
\end{align}
Since we have already known these expectation values (\ref{Ex:psi:L}), (\ref{Ex:psi:R}) and (\ref{ExComRel:JF:L}), (\ref{ExComRel:JF:R}), substituting them we obtain  
\begin{align}
\langle\psi^{\alpha}_{1/2}\tilde{G}^{\rm M}_{1/2}e^{+}_{\mu\nu}\tilde{\psi}^{\mu}_{-1/2}\psi^{\nu}_{-1/2}\rangle
&=-i\sqrt{\frac{\alpha'}{2}}\Big[\partial_{\lambda}e^{+}_{\mu\nu}g^{\lambda\mu}
g^{\alpha\nu}+e^{+}_{\mu\nu}(f^{\mu\nu}g^{\alpha+}+k^{\mu\nu}_{1}g^{\alpha z}
+k^{\mu\nu}_{2}g^{\alpha z*})\Big],\\
\langle\tilde{\psi}^{\alpha}_{1/2}G^{\rm M}_{1/2}e^{+}_{\mu\nu}\tilde{\psi}^{\mu}_{-1/2}\psi^{\nu}_{-1/2}\rangle
&=i\sqrt{\frac{\alpha'}{2}}\left[\partial_{\lambda}e^{+}_{\mu\nu}g^{\alpha\mu}g^{\lambda\nu}+e^{+}_{\mu\nu}(\tilde{f}^{\mu\nu}g^{\alpha+}
+k_{1}^{\mu\nu}g^{\alpha z}+k_{2}^{\mu\nu}g^{\alpha z*})\right],
\end{align}
where $\langle\tilde{\psi}^{\alpha}_{1/2}G^{\rm M}_{1/2}e^{+}_{\mu\nu}\tilde{\psi}^{\mu}_{-1/2}\psi^{\nu}_{-1/2}\rangle$ is obtained by similar calculation.
Here we define $f^{\mu\nu}$, $\tilde{f}^{\mu\nu}$, $k^{\mu\nu}_{1}$ and $k^{\mu\nu}_{2}$ respectively as 
\begin{align}
f^{\mu\nu}&=\frac{i\mu}{2}\left[g^{z[\mu}g^{\nu]z*}+i\mu(zg^{+[\mu}g^{\nu]z*}
+z^{*}g^{+[\mu}g^{\nu]z})\right],\label{f:munu}\\
\tilde{f}^{\mu\nu}&=\frac{i\mu}{2}\left[g^{z[\mu}g^{\nu]z*}-i\mu(zg^{+[\mu}g^{\nu]z*}
+z^{*}g^{+[\mu}g^{\nu]z})\right],\label{tilde:f:munu}\\
k^{\mu\nu}_{1}&=\frac{1}{2}\left(\mu^{2}z^{*}g^{\mu+}g^{\nu+}-i\mu g^{+[\mu}g^{\nu]z*}\right),\ \ 
k^{\mu\nu}_{2}=\frac{1}{2}\left(\mu^{2}z g^{\mu+}g^{\nu+}+i\mu g^{+[\mu}g^{\nu]z}\right)\label{k:munu}.
\end{align}
We have to note that $f^{\mu\nu}$ and $\tilde{f}^{\mu\nu}$ are antisymmetric tensor so that they vanish in the case of the product with symmetric tensor.
Since we obtain  expectation values related to the fermionic super-Virasoro operator, we can calculate the terms in $\bar{B}_{\alpha}\bar{B}^{\alpha}$ and $\bar{E}_{\alpha}\bar{E}^{\alpha}$. First the term in $\bar{B}_{\alpha}\bar{B}^{\alpha}$ is
\begin{align}
g_{\alpha\beta}&\langle\psi^{\alpha}_{1/2}\tilde{G}^{\rm M}_{1/2}e^{+}_{\mu\nu}
\tilde{\psi}^{\mu}_{-1/2}\psi^{\nu}_{-1/2}\rangle
\langle\psi^{\beta}_{1/2}\tilde{G}^{\rm M}_{1/2}e^{+}_{\rho\sigma}
\tilde{\psi}^{\rho}_{-1/2}\psi^{\sigma}_{-1/2}\rangle\nonumber\\
&=-\frac{\alpha'}{2}\Big[
\partial_{\lambda}e^{+}_{\mu\nu}\partial_{\omega}e^{+}_{\rho\sigma}g^{\lambda\mu}g^{\omega\rho}g^{\nu\sigma}+2e^{+}_{\rho\sigma}\partial_{\lambda}e^{+}_{\mu\nu}g^{\lambda\mu}(f^{\rho\sigma}g^{\nu+}+k^{\rho\sigma}_{1}g^{\nu z}+k^{\rho\sigma}_{2}g^{\nu z*})+4e^{+}_{\mu\nu}e^{+}_{\rho\sigma}k^{\mu\nu}_{1}
k^{\rho\sigma}_{2}\Big]\label{BB}.
\end{align}
Next the terms in $\bar{E}_{\alpha}\bar{E}^{\alpha}$ are
\begin{align}
g_{\alpha\beta}&\langle\tilde{\psi}^{\alpha}_{1/2}G^{\rm M}_{1/2}e^{+}_{\mu\nu}
\tilde{\psi}^{\mu}_{-1/2}\psi^{\nu}_{-1/2}\rangle
\langle\tilde{\psi}^{\beta}_{1/2}G^{\rm M}_{1/2}e^{+}_{\rho\sigma}
\tilde{\psi}^{\rho}_{-1/2}\psi^{\sigma}_{-1/2}\rangle\nonumber\\
&=-\frac{\alpha'}{2}\left[
\partial_{\lambda}e^{+}_{\mu\nu}\partial_{\omega}e^{+}_{\rho\sigma}g^{\mu\rho}g^{\lambda\nu}g^{\omega\sigma}
+2e^{+}_{\rho\sigma}\partial_{\lambda}e^{+}_{\mu\nu}g^{\lambda\nu}(\tilde{f}^{\rho\sigma}g^{\mu+}+k^{\rho\sigma}_{1}g^{\mu z}+k^{\rho\sigma}_{2}g^{\mu z*})+4e^{+}_{\mu\nu}e^{+}_{\rho\sigma}k^{\mu\nu}_{1}
k^{\rho\sigma}_{2}\right],\label{EE1}\\
g_{\alpha\beta}&\langle\tilde{\psi}^{\alpha}_{1/2}G^{\rm M}_{1/2}e^{+}_{\mu\nu}
\tilde{\psi}^{\mu}_{-1/2}\psi^{\nu}_{-1/2}\rangle
\langle\tilde{\psi}^{\beta}_{1/2}\tilde{G}^{\rm M}_{-1/2}\phi\rangle\nonumber\\
&=\frac{\alpha'}{2}\left[
\partial_{\lambda}e^{+}_{\mu\nu}g^{\omega\mu}g^{\lambda\nu}
+e^{+}_{\mu\nu}(\tilde{f}^{\mu\nu}g^{\omega+}+k^{\mu\nu}_{1}g^{\omega z}+k^{\mu\nu}_{2}g^{\omega z*})\right]\partial_{\omega}\phi,\label{EE2}\\
g_{\alpha\beta}&\langle\tilde{\psi}^{\alpha}_{1/2}\tilde{G}^{\rm M}_{-1/2}\phi\rangle
\langle\tilde{\psi}^{\beta}_{1/2}\tilde{G}^{\rm M}_{-1/2}\phi\rangle
=-\frac{\alpha'}{2}g^{\lambda\omega}\partial_{\lambda}\phi\partial_{\omega}\phi\label{EE3}.
\end{align}
In the previous subsection although we calculate the bosonic super-Virasoro operator part of the action, we do not find the term like $\partial_{\lambda}e^{+}_{\mu\nu}\partial_{\omega}e^{+}_{\rho\sigma}$ in (\ref{BB}) and (\ref{EE1}), so that we had better make the term like $e^{+}_{\mu\nu}\partial_{\lambda}\partial_{\omega}e^{+}_{\rho\sigma}$ using the identity,
\begin{align}
g^{\lambda\mu}g^{\omega\rho}g^{\nu\sigma}\partial_{\lambda}e^{+}_{\mu\nu}\partial_{\omega}e^{+}_{\rho\sigma}
=\Big[\partial_{\lambda}(g^{\lambda\mu}g^{\omega\rho}g^{\nu\sigma}e^{+}_{\mu\nu}\partial_{\omega}
e^{+}_{\rho\sigma})-e^{+}_{\mu\nu}\partial_{\lambda}(g^{\lambda\mu}g^{\omega\rho}g^{\nu\sigma}\partial_{\omega}
e^{+}_{\rho\sigma})\Big].
\end{align}
Using the property of the inverse of the pp-wave metric (\ref{der:metric}),
the above identity becomes
\begin{align}
 &g^{\lambda\mu}g^{\omega\rho}g^{\nu\sigma}\partial_{\lambda}e^{+}_{\mu\nu}\partial_{\omega}e^{+}_{\rho\sigma}
\nonumber\\
&=-\Big[g^{\lambda\mu}g^{\omega\rho}g^{\nu\sigma}e^{+}_{\mu\nu}\partial_{\lambda}\partial_{\omega}e^{+}_{\rho\sigma}+2{\mu}^{2}g^{\nu\sigma}(ze^{+}_{z\nu}\partial_{-}e^{+}_{-\sigma}+z^{*}e^{+}_{z*\nu}\partial_{-}e^{+}_{-\sigma})
+2{\mu}^{2}g^{\omega\rho}(ze^{+}_{z-}\partial_{\omega}e^{+}_{\rho-}+z^{*}e^{+}_{z*-}\partial_{\omega}e^{+}_{\rho-})\Big]
\nonumber\\
&\ \ \ \ +\partial_{\lambda}(g^{\lambda\mu}g^{\omega\rho}g^{\nu\sigma}e^{+}_{\mu\nu}\partial_{\omega}
e^{+}_{\rho\sigma}).
\end{align} 
The last term of this identity is not important because we can integrate out this total derivative term. Next we evaluate the second term of (\ref{BB}). Because of the symmetry of spacetime about the coefficient $e^{+}_{\rho\sigma}\partial_{\lambda}e^{+}_{\mu\nu}$ and $f^{\mu\nu}$, $k^{\mu\nu}_{1}$ and $k^{\mu\nu}_{2}$ which are made of the pp-wave metric (\ref{f:munu})-(\ref{k:munu}), the second term of (\ref{BB}) have already been determined then we obtain
\begin{align}
2e^{+}_{\rho\sigma}&\partial_{\lambda}e^{+}_{\mu\nu}g^{\lambda\mu}(f^{\rho\sigma}g^{\nu+}+k^{\rho\sigma}_{1}g^{\nu z}+k^{\rho\sigma}_{2}g^{\nu z*})\nonumber\\
&=-8i\mu b_{z*z}g^{\lambda\mu}\partial_{\lambda}e^{+}_{\mu-}
-4\mu^{2}(zb_{-z}+z^{*}b_{-z*})g^{\lambda\mu}\partial_{\lambda}e^{+}_{\mu-}
\nonumber\\
&\ \ \ \ +2\mu^{2}h_{--}g^{\lambda\mu}(z\partial_{\lambda}e^{+}_{\mu z}+z^{*}\partial_{\lambda}e^{+}_{\mu z*})
+8i\mu g^{\lambda\mu}(b_{-z}\partial_{\lambda}e^{+}_{\mu z*}-b_{-z*}\partial_{\lambda}e^{+}_{\mu z}).
\end{align}
Similarly we can evaluate the term of $e^{+}_{\rho\sigma}\partial_{\lambda}e^{+}_{\mu\nu}$ in the second term of (\ref{EE1}) as
\begin{align}
2e^{+}_{\rho\sigma}&\partial_{\lambda}e^{+}_{\mu\nu}g^{\lambda\nu}(\tilde{f}^{\rho\sigma}g^{\mu+}+k^{\rho\sigma}_{1}g^{\mu z}+k^{\rho\sigma}_{2}g^{\mu z*})\nonumber\\
&=-8i\mu b_{z*z}g^{\lambda\nu}\partial_{\lambda}e^{+}_{-\nu}
+4\mu^{2}(zb_{-z}+z^{*}b_{-z*})g^{\lambda\nu}\partial_{\lambda}e^{+}_{-\nu}
\nonumber\\
&\ \ \ \ +2\mu^{2}h_{--}g^{\lambda\nu}(z\partial_{\lambda}e^{+}_{z\nu}+z^{*}\partial_{\lambda}e^{+}_{z*\nu})
+8i\mu g^{\lambda\nu}(b_{-z}\partial_{\lambda}e^{+}_{z*\nu}-b_{-z*}\partial_{\lambda}e^{+}_{z\nu}).
\end{align}
Moreover the term of $e^{+}_{\mu\nu}e^{+}_{\rho\sigma}$ which is common in both (\ref{BB}) and (\ref{EE1}) is
\begin{align}
4e^{+}_{\mu\nu}&e^{+}_{\rho\sigma}k^{\mu\nu}_{1}k^{\rho\sigma}_{2}
=\mu^{4}z^{*}zh_{--}h_{--}+4i\mu^{3}h_{--}(zb_{-z}-z^{*}b_{-z*})
+16\mu^{2}b_{-z}b_{-z*}.
\end{align}
Next we evaluate the term of $e^{+}_{\mu\nu}$ and scalar field$\phi$ in (\ref{EE2}). Taking the partial integral, the first term in (\ref{EE2}) becomes
\begin{align}
-2\partial_{\lambda}e^{+}_{\mu\nu}g^{\omega\mu}g^{\lambda\nu}\partial_{\omega}
\phi
=2\phi g^{\omega\mu}g^{\lambda\nu}\partial_{\omega}\partial_{\lambda}e^{+}_{\mu\nu}+4\mu^{2}\phi(z\partial_{-}e^{+}_{z-}+z^{*}\partial_{-}e^{+}_{z*-})
-\partial_{\omega}(2\partial_{\lambda}e^{+}_{\mu\nu}g^{\omega\mu}g^{\lambda\nu}
\phi).
\end{align}
In the calculation of the partial integral for the second term in (\ref{EE2}), we use the formulae as 
\begin{align}
\partial_{\omega}f^{\mu\nu}&=-\frac{\mu^{2}}{2}(\delta^{z}_{\omega}g^{+[\mu}g^{\nu]z*}+\delta^{z*}_{\omega}g^{+[\mu}g^{\nu]z}),\ \ 
\partial_{\omega}\tilde{f}^{\mu\nu}=\frac{\mu^{2}}{2}(\delta^{z}_{\omega}g^{+[\mu}g^{\nu]z*}+\delta^{z*}_{\omega}g^{+[\mu}g^{\nu]z}),\label{d:f:munu}\\
\partial_{\omega}k_{1}^{\mu\nu}&=\frac{\mu^{2}}{2}\delta^{z*}_{\omega}g^{\mu+}g^{\nu+},\ \ \ \ \ \ \ \ \ \ \ \ \ \ \ \ \ \ \ \ \ \ \ \ \ \ 
\partial_{\omega}k_{2}^{\mu\nu}=\frac{\mu^{2}}{2}\delta^{z}_{\omega}g^{\mu+}g^{\nu+}\label{d:k:munu}.
\end{align}
which are obtained by the differenciating (\ref{f:munu})-(\ref{k:munu}).
Using the formulae of (\ref{d:f:munu}) and (\ref{d:k:munu}), we can calculate the second term in (\ref{EE2}) as 
\begin{align}
e^{+}_{\mu\nu}&(\tilde{f}^{\mu\nu}g^{\omega+}+k^{\mu\nu}_{1}g^{\omega z}
+k^{\mu\nu}_{2}g^{\omega z*})\partial_{\omega}\phi\nonumber\\
&=-\phi\partial_{\omega}\left[e^{+}_{\mu\nu}(\tilde{f}^{\mu\nu}g^{\omega+}+k^{\mu\nu}_{1}g^{\omega z}+k^{\mu\nu}_{2}g^{\omega z*})\right]
+\partial_{\omega}\left[e^{+}_{\mu\nu}(\tilde{f}^{\mu\nu}g^{\omega+}+k^{\mu\nu}_{1}g^{\omega z}+k^{\mu\nu}_{2}g^{\omega z*})\phi\right]\nonumber\\
&=-\phi\Big[4i\mu(\partial_{-}b_{zz*}+\partial_{z}b_{z*-}+\partial_{z*}b_{-z})+2\mu^{2}(z\partial_{-}b_{-z}+z^{*}\partial_{-}b_{-z*})+\mu^{2}(z\partial_{z}h_{--}+z^{*}\partial_{z*}h_{--})\Big]\nonumber\\
&\ \ \ \ \ \ \ \ \ \ -2\mu^{2}\phi h_{--}+({\rm total\ derivative}).
\end{align}
Therefore gathering all the terms, we can obtain the action $S_{2}$:
\begin{align}
S_{2}=-\frac{\alpha'}{4}\int d^{10}x\Big\{&-2g^{\lambda\mu}g^{\omega\rho}g^{\nu\sigma}h_{\mu\nu}\partial_{\lambda}\partial_{\omega}h_{\rho\sigma}-2g^{\lambda\mu}g^{\omega\rho}g^{\nu\sigma}b_{\mu\nu}\partial_{\lambda}
\partial_{\omega}b_{\rho\sigma}\nonumber\\
&+4\mu^{2}g^{\mu\nu}\Big[h_{-\mu}(z\partial_{-}h_{z\nu}+z^{*}\partial_{-}h_{z*\nu})
+b_{-\mu}(z\partial_{-}b_{z\nu}+z^{*}\partial_{-}b_{z*\nu})\Big]\nonumber\\
&-4\mu^{2}g^{\mu\nu}\Big[(zh_{-z}+z^{*}h_{-z*})\partial_{\mu}h_{\nu-}+(zb_{-z}+z^{*}b_{-z*})\partial_{\mu}b_{\nu-}\Big]
\nonumber\\
&+16i\mu g^{\lambda\mu}\Big[b_{-z}\partial_{\lambda}h_{\mu z*}-b_{-z*}\partial_{\lambda}h_{\mu z}+b_{zz*}\partial_{\lambda}h_{\mu-}\Big]\nonumber\\
&+4\mu^{2}h_{--}g^{\lambda\mu}(z\partial_{\lambda}h_{\mu z}+z^{*}\partial_{\lambda}h_{\mu z*})\nonumber\\
&+2\mu^{4}z^{*}zh_{--}h_{--}+8i\mu^{3}h_{--}(zb_{-z}-z^{*}b_{-z*})+32\mu^{2}b_{-z}b_{-z*}
\nonumber\\
&+g^{\lambda\omega}\partial_{\lambda}\phi\partial_{\omega}\phi+2\phi g^{\omega\mu}g^{\lambda\nu}\partial_{\omega}\partial_{\lambda}h_{\mu\nu}
+2\phi g^{\omega\mu}g^{\lambda\nu}\partial_{\omega}\partial_{\lambda}b_{\mu\nu}\nonumber\\
&+4\mu^{2}\phi(z\partial_{-}h_{-z}+z^{*}\partial_{-}h_{-z*})+2\mu^{2}\phi(z\partial_{z}h_{--}+z^{*}\partial_{z*}h_{--})
+4\mu^{2}\phi h_{--}\nonumber\\
&+8i\mu\phi(\partial_{-}b_{zz*}+\partial_{z}b_{z*-}+\partial_{z*}b_{-z})
\Big\}.\label{S2:final}
\end{align}
The component representation of superstring field action is the sum of (\ref{S1:final}) and (\ref{S2:final}) namely $S=S_{1}+S_{2}$.
In this calculation the term $8i\mu^{3}h_{--}(zb_{-z}-z^{*}b_{-z*})$ vanishes trivially and performing the partial integral of the term $-4\mu^{2}g^{\mu\nu}(zb_{-z}+z^{*}b_{-z*})\partial_{\mu}b_{\nu-}$ in (\ref{S2:final}), it becomes $-16\mu^{2}b_{-z}b_{-z*}+4\mu^{2}g^{\mu\nu}b_{\nu-}(z\partial_{\mu}b_{-z}+z^{*}\partial_{\mu}b_{-z*})+({\rm total\ derivative})$, therefore $-16\mu^{2}b_{-z}b_{-z*}$ coming from this, $32\mu^{2}b_{-z}b_{-z*}$ in (\ref{S2:final}) and $-16\mu^{2}b_{-z}b_{-z*}$ in (\ref{S1:final}) are canceled out each other. Moreover we can gather the same kind of terms using the other partial integrals. The total action becomes 
\begin{align}
S=-\frac{\alpha'}{4}\int d^{10}x\Big\{&h^{\mu\nu}g^{\rho\sigma}\partial_{\rho}\partial_{\sigma}h_{\mu\nu}-2g^{\lambda\mu}g^{\omega\rho}g^{\nu\sigma}h_{\mu\nu}\partial_{\lambda}\partial_{\omega}h_{\rho\sigma}\nonumber\\
&-4\mu^{2}g^{\mu\nu}h_{-\mu}(z\partial_{-}h_{z\nu}+z^{*}\partial_{-}h_{z*\nu})
-4\mu^{2}h^{+\mu}(z\partial_{z}h_{-\mu}+z^{*}\partial_{z*}h_{-\mu})\nonumber\\
&-4\mu^{2}g^{\mu\nu}(zh_{-z}+z^{*}h_{-z*})\partial_{\mu}h_{\nu-}
+4\mu^{2}h_{--}g^{\lambda\mu}(z\partial_{\lambda}h_{\mu z}+z^{*}\partial_{\lambda}h_{\mu z*})\nonumber\\
&-4\mu^{2}(h^{+z}h_{-z}+h^{+z*}h_{-z*})+4\mu^{2}z^{*}zh^{++}h_{--}-16\mu^{2}h^{++}h_{zz*}\nonumber\\
&-g^{\rho\sigma}g^{\alpha\nu}g^{\beta\mu}b_{\alpha\beta}\partial_{\rho}\partial_{\sigma}b_{\mu\nu}
-2g^{\lambda\mu}g^{\omega\rho}g^{\nu\sigma}b_{\mu\nu}\partial_{\lambda}\partial_{\omega}b_{\rho\sigma}\nonumber\\
&+4\mu^{2}g^{\beta\nu}b_{-\beta}(z\partial_{z}b_{-\nu}+z^{*}\partial_{z*}b_{-\nu})
-4\mu^{2}g^{\beta\nu}b_{-\beta}(z\partial_{-}b_{z\nu}+z^{*}\partial_{-}b_{z*\nu})\nonumber\\
&+4\mu^{2}g^{\mu\nu}b_{\nu-}(z\partial_{\mu}b_{-z}+z^{*}\partial_{\mu}b_{-z*})\nonumber\\
&-16i\mu g^{\beta\nu}\Big[h_{z*\beta}(\partial_{\nu}b_{-z}+\partial_{-}b_{z\nu}-\partial_{z}b_{\nu-})
-h_{z\beta}(\partial_{\nu}b_{-z*}+\partial_{-}b_{z*\nu}-\partial_{z*}b_{-\nu})\nonumber\\
&\ \ \ \ \ \ \ \ \ \ \ \ \ \ \ +h_{-\beta}(\partial_{\nu}b_{zz*}+\partial_{z}b_{z*\nu}-\partial_{z*}b_{z\nu})\Big]
\nonumber\\
&+g^{\lambda\omega}\partial_{\lambda}\phi\partial_{\omega}\phi+2\phi g^{\omega\mu}g^{\lambda\nu}\partial_{\omega}
\partial_{\lambda}h_{\mu\nu}\nonumber\\
&+4\mu^{2}\phi(z\partial_{-}h_{-z}+z^{*}\partial_{-}h_{-z*})+2\mu^{2}\phi(z\partial_{z}h_{--}+z^{*}\partial_{z*}h_{--})
+4\mu^{2}\phi h_{--}\nonumber\\
&+2\phi g^{\omega\mu}g^{\lambda\nu}\partial_{\omega}\partial_{\lambda}b_{\mu\nu}
+8i\mu\phi(\partial_{-}b_{zz*}+\partial_{z}b_{z*-}+\partial_{z*}b_{-z})
\Big\}.
\end{align}
Since the trace part of $h_{\mu\nu}$ is absent in this action so that we replace $\phi\rightarrow h-4\phi$ where $h$ is trace part of $h_{\mu\nu}$ namely $h=g^{\mu\nu}h_{\mu\nu}$. Finally the action becomes as follows:
\begin{align}
S=-\frac{\alpha'}{4}\int d^{10}x\Big\{&g^{\rho\sigma}h^{\mu\nu}(\partial_{\rho}\partial_{\sigma}h_{\mu\nu}-2\partial_{\rho}\partial_{\nu}h_{\mu\sigma})+2h^{\mu\nu}\partial_{\mu}\partial_{\nu}h-g^{\rho\sigma}h\partial_{\rho}\partial_{\sigma}h\nonumber\\
&+4\mu^{2}h^{+\mu}(z\partial_{-}h_{\mu z}+z^{*}\partial_{-}h_{\mu z*})-4\mu^{2}h^{+\mu}(z\partial_{z}h_{-\mu}+z^{*}\partial_{z*}h_{-\mu})\nonumber\\
&+2\mu^{2}(z^{*}h^{+z}+zh^{+z*})g^{\rho\sigma}\partial_{\rho}h_{-\sigma}
+4\mu^{2}h^{++}g^{\rho\sigma}(z\partial_{\rho}h_{\sigma z}+z^{*}\partial_{\rho}h_{\sigma z*})\nonumber\\
&-4\mu^{2}(h^{+z}h_{-z}+h^{+z*}h_{-z*})+4\mu^{2}z^{*}zh^{++}h_{--}-16\mu^{2}h^{++}h_{zz*}\nonumber\\
&-2\mu^{2}(zh^{+z*}+z^{*}h^{+z})\partial_{-}h-2\mu^{2}h^{++}(z\partial_{z}h+z^{*}\partial_{z*}h)\nonumber\\
&+g^{\mu\alpha}g^{\nu\beta}g^{\rho\gamma}b_{\nu\rho}(\partial_{\mu}\partial_{\alpha}b_{\beta\gamma}-2\partial_{\mu}\partial_{\beta}b_{\alpha\gamma})+4\mu^{2}g^{\rho\gamma}b_{-\rho}(z\partial_{z}b_{-\gamma}
+z^{*}\partial_{z*}b_{-\gamma})\nonumber\\
&-4\mu^{2}g^{\rho\gamma}b_{-\rho}(z\partial_{-}b_{z\gamma}+z^{*}\partial_{-}b_{z*\gamma})
+4\mu^{2}g^{\nu\beta}b_{\nu-}(z\partial_{\beta}b_{-z}+z^{*}\partial_{\beta}b_{-z*})\nonumber\\
&-16i\mu g^{\mu\nu}\Big[h_{z*\mu}(\partial_{\nu}b_{-z}+\partial_{-}b_{z\nu}-\partial_{z}b_{\nu-})
-h_{z\mu}(\partial_{\nu}b_{-z*}+\partial_{-}b_{z*\nu}-\partial_{z*}b_{-\nu})\nonumber\\
&+h_{-\mu}(\partial_{\nu}b_{zz*}+\partial_{z}b_{z*\nu}-\partial_{z*}b_{z\nu})
-\frac{1}{2}h_{\mu\nu}(\partial_{-}b_{zz*}+\partial_{z}b_{z*-}+\partial_{z*}b_{-z})\Big]\nonumber\\
&+16g^{\mu\nu}\partial_{\mu}\phi\partial_{\nu}\phi+8g^{\mu\nu}\phi\partial_{\mu}\partial_{\nu}h
-8\phi g^{\rho\mu}g^{\nu\sigma}\partial_{\rho}\partial_{\sigma}h_{\mu\nu}-16\mu^{2}\phi h_{--}\nonumber\\
&-16\mu^{2}\phi(z\partial_{-}h_{-z}+z^{*}\partial_{-}h_{-z*})-8\mu^{2}\phi(z\partial_{z}h_{--}+z^{*}\partial_{z*}h_{--})
\nonumber\\
&-32i\mu\phi(\partial_{-}b_{zz*}+\partial_{z}b_{z*-}+\partial_{z*}b_{-z})
\Big\}.\label{FinalAction2}
\end{align}
Since we treat ``covariant''string field action, it is important that the spacetime light-cone directions $\mu,\nu=\pm$ remain in this action. If we treat the light-cone string field action, all the term containing the light-cone components vanish and we cannot see them. In the next section we compare this action to the second-order perturbation of supergravity action then we see the exact correspondence between them. It is important that we find this corrrespondence except for the flat background. Of course in the limit of $\mu\rightarrow 0$, we can reproduse the action in the case of the flat background. 


\section{Comparison with the second-order perturbation of supergravity action}
In this section we compare the NS-NS sector of the low energy superstring field action with the second-order perturbation of a  part of supergravity action in the NS-NS pp-wave background. Because of being compared to NS-NS sector of superstring field action, here, we do not consider R-R fields and any fermions such as gravitino and dilatino. The supergravity action of the part of gravitational field, NS-NS anti-symmetric field and dilaton field is
\begin{align}
S=-\frac{1}{2\kappa^{2}}\int d^{10}x\sqrt{-G}e^{-2\hat{\Phi}}
\left[R+4G^{\mu\nu}\partial_{\mu}\hat{\Phi}\partial_{\nu}\hat{\Phi}
-\frac{1}{12}G^{\alpha\mu}G^{\beta\nu}G^{\gamma\rho}
H_{\alpha\beta\gamma}H_{\mu\nu\rho}\right].
\end{align}
Here because of avoidance of confusion between string field and dilaton field, we use $\hat{\Phi}$ for the dilaton.
For caluculation of the second-order perturbation of this action, we define perturbation of fields as
\begin{align}
G_{\mu\nu}=g_{\mu\nu}+h_{\mu\nu},\ B_{\mu\nu}=\bar{B}_{\mu\nu}+b_{\mu\nu},\ 
\hat{\Phi}=\bar{\Phi}+\phi.
\end{align}
Here $g_{\mu\nu}$, $\bar{B}_{\mu\nu}$ and $\bar{\Phi}$ denote background gravitational field, background NS-NS antisymmetric tensor field and background dilaton field respectively moreover $h_{\mu\nu}$, $b_{\mu\nu}$ and $\phi$ denote perturbative gravitational field, purturbative NS-NS antisymmetric tensor field and purturbative dilaton field respectively. We will find that $h_{\mu\nu}$, $b_{\mu\nu}$ and $\phi$ correspond to coefficients of component field expansions of string field after calculation of the second-order perturbation of supergravity action. In the perturbation contravariant metric $G^{\mu\nu}$, $\sqrt{-G}$ and $e^{-2\hat{\Phi}}$ become 
\begin{align}
G^{\mu\nu}&=g^{\mu\nu}-h^{\mu\nu}+h^{\mu\alpha}h_{\alpha}^{\nu}+{\cal O}(h^{3}),\\
\sqrt{-G}&=\sqrt{-g}\left\{1+\frac{1}{2}h
+\frac{1}{8}(h^{2}-2g^{\mu\alpha}g^{\nu\beta}h_{\mu\nu}h_{\alpha\beta})\right\}+{\cal O}(h^{3}),\\
e^{-2\hat{\Phi}}&=e^{-2\bar{\Phi}}(1-2\phi+2\phi^{2})+{\cal O}(\phi^{3}),
\end{align}
up to second-order perturbation. Moreover Ricci scalar $R$ is perturbed up to the second-order as 
\begin{align}
R&=R^{(0)}+R^{(1)}+R^{(2)}+{\cal O}(h^{3}),\\
R^{(1)}&=-\nabla^{\alpha}\nabla_{\alpha}h+\nabla^{\alpha}\nabla_{\beta}
h_{\alpha}^{\beta}-R^{(0)}_{\alpha\beta}h^{\alpha\beta},\\
R^{(2)}&=\frac{1}{4}h^{\mu\nu}\nabla_{\alpha}\nabla^{\alpha}h_{\mu\nu}
-\frac{1}{2}h^{\mu\beta}\nabla^{\alpha}\nabla_{\beta}h_{\mu\alpha}
+\frac{1}{4}h\nabla_{\alpha}\nabla^{\alpha}h+R^{(0)}_{\alpha\beta}h^{\alpha\omega}h_{\omega}^{\beta}.
\end{align}
Here $R^{(0)}$, $R^{(1)}$ and $R^{(2)} $denote background Ricci scalar, the first order perturbation of Ricci scalar and the second-order perturbation of Ricci scalar respectively and $R^{(0)}_{\alpha\beta}$ denotes the background Ricci tensor. As is trivial, the flux of the NS-NS antisymmetric field $H_{\mu\nu\rho}$ is perturbed as 
\begin{align}
H_{\mu\nu\rho}&=\bar{H}_{\mu\nu\rho}+H^{(1)}_{\mu\nu\rho},\\
\bar{H}_{\mu\nu\rho}&=\partial_{\mu}\bar{B}_{\nu\rho}+\partial_{\nu}\bar{B}_{\rho\mu}
+\partial_{\rho}\bar{B}_{\mu\nu},\\
H^{(1)}_{\mu\nu\rho}&=\partial_{\mu}b_{\nu\rho}+\partial_{\nu}b_{\rho\mu}
+\partial_{\rho}b_{\mu\nu}.
\end{align}
In general, using these perturbation, we can obtain the following general second-order action in some background fields:
\begin{align}
S=-\frac{1}{2\kappa^{2}}\int d^{10}x\sqrt{-g}\Big\{e^{-2\bar{\Phi}}
\Big[R^{(2)}+4g^{\mu\nu}\partial_{\mu}\phi\partial_{\nu}\phi
-8h^{\mu\nu}\partial_{\mu}\bar{\Phi}\partial_{\nu}\phi
+(4\partial_{\mu}\bar{\Phi}\partial_{\nu}\bar{\Phi}
-\frac{1}{4}\bar{H}_{\mu\alpha\beta}\bar{H}_{\nu}^{\ \alpha\beta})h^{\mu\omega}h_{\omega}^{\nu}\nonumber\\
-\frac{1}{4}\bar{H}^{\alpha\beta}_{\ \ \gamma}\bar{H}^{\mu\nu\gamma}h_{\alpha\mu}h_{\beta\nu}
-\frac{1}{12}g^{\alpha\mu}g^{\beta\nu}g^{\gamma\rho}H^{(1)}_{\alpha\beta\gamma}
H^{(1)}_{\mu\nu\rho}+g^{\mu\nu}\bar{H}^{\alpha\beta\gamma}H^{(1)}_{\mu\beta\gamma}h_{\alpha\nu}\Big]\nonumber\\
+\frac{1}{2}(h-4\phi)\Big[
R^{(1)}
-\frac{1}{6}\bar{H}^{\mu\nu\rho}H^{(1)}_{\mu\nu\rho}
+8g^{\mu\nu}\partial_{\mu}\bar{\Phi}\partial_{\nu}\phi
-(4\partial_{\mu}\bar{\Phi}\partial_{\nu}\bar{\Phi}
-\frac{1}{4}\bar{H}_{\mu\alpha\beta}\bar{H}_{\nu}^{\ \alpha\beta})h^{\mu\nu}\Big]
\Big\}\label{GPA},
\end{align}
where using the background field equations, the zeroth-order and the first-order perturbation of the action vanish and we also remove trivially vanishing terms in the second-order perturbation. Especially using the Einstein equation $R^{(0)}_{\mu\nu}-\frac{1}{4}\bar{H}_{\mu\alpha\beta}\bar{H}_{\nu}^{\alpha\beta}+2\nabla_{\mu}\partial_{\nu}\bar{\Phi}=0$, we can remove background Ricci tensor $R^{(0)}_{\mu\nu}$ in $R^{(1)}$ and $R^{(2)}$ from general action (\ref{GPA}). 
Moreover in the case of $\bar{\Phi}=0$ like the NS-NS pp-wave background, this action becomes the simplest form and we had better take partial integral in the term of $H^{(1)}_{\mu\nu\rho}$ for comparing superstring field action in the previous section.
The action becomes 
\begin{align}
S=-\frac{1}{2\kappa^{2}}\int d^{10}x\sqrt{-g}
\bigg[\frac{1}{4}h^{\mu\nu}(\nabla_{\rho}\nabla^{\rho}h_{\mu\nu}
-2\nabla^{\rho}\nabla_{\nu}h_{\mu\rho}
+2\nabla_{\mu}\nabla_{\nu}h
-g_{\mu\nu}\nabla_{\rho}\nabla^{\rho}h)\nonumber\\
-\frac{1}{4}h^{\alpha\mu}h^{\beta\nu}
\bar{H}_{\alpha\beta\gamma}\bar{H}_{\mu\nu}^{\ \ \gamma}
+\frac{1}{4}g^{\mu\alpha}g^{\nu\beta}g^{\rho\gamma}
b_{\nu\beta}\nabla_{\mu}(\partial_{\alpha}b_{\beta\gamma}
-2\partial_{\beta}b_{\alpha\gamma})\nonumber\\
+\frac{1}{2}g^{\mu\nu}\bar{H}^{\alpha\beta\gamma}
h_{\gamma\mu}(\partial_{\nu}b_{\alpha\beta}
+2\partial_{\alpha}b_{\beta\nu})
-\frac{1}{4}\bar{H}^{\mu\nu\rho}\partial_{\mu}b_{\nu\rho}h\nonumber\\
+4g^{\mu\nu}\partial_{\mu}\phi\partial_{\nu}\phi
+\bar{H}^{\mu\nu\rho}\partial_{\mu}b_{\nu\rho}\phi
+2\phi(\nabla_{\mu}\nabla^{\mu}h-\nabla^{\mu}\nabla^{\nu}h_{\mu\nu})
\bigg].
\end{align}
After this, we substitute the NS-NS pp-wave background into the general second-order perturbation of supergravity. In the case of the NS-NS pp-wave background (\ref{pp-wave:metric}) and (\ref{NS-NS:flux}), $\sqrt{-g}=\frac{1}{4}$ and non-vanishing components of the Christoffel symbol and the field strength of NS-NS antisymmetric tensor are 
\begin{align}
\bar{\Gamma}^{-}_{+z}&=\bar\Gamma^{-}_{z+}=\frac{1}{2}\mu^{2}z^{*},\ 
\bar{\Gamma}^{-}_{+z*}=\bar\Gamma^{-}_{z*+}=\frac{1}{2}\mu^{2}z,\\ \bar{\Gamma}^{z}_{++}&=\mu^{2}z,\ \bar{\Gamma}^{z*}_{++}=\mu^{2}z^{*},\\
 \bar{H}_{+zz*}&=-i\mu.
\end{align}
Using these Chistoffel symbol and the field strength of NS-NS field, we can calculate the covariant derivative so that we can represent the action by using components. Finally the action becomes the following form:
\begin{align}
S=-\frac{1}{32\kappa^{2}}\int d^{10}x\Big\{&g^{\rho\sigma}h^{\mu\nu}(\partial_{\rho}\partial_{\sigma}h_{\mu\nu}-2\partial_{\rho}\partial_{\nu}h_{\mu\sigma})+2h^{\mu\nu}\partial_{\mu}\partial_{\nu}h-g^{\rho\sigma}h\partial_{\rho}\partial_{\sigma}h\nonumber\\
&+4\mu^{2}h^{+\mu}(z\partial_{-}h_{\mu z}+z^{*}\partial_{-}h_{\mu z*})-4\mu^{2}h^{+\mu}(z\partial_{z}h_{-\mu}+z^{*}\partial_{z*}h_{-\mu})\nonumber\\
&+2\mu^{2}(z^{*}h^{+z}+zh^{+z*})g^{\rho\sigma}\partial_{\rho}h_{-\sigma}
+4\mu^{2}h^{++}g^{\rho\sigma}(z\partial_{\rho}h_{\sigma z}+z^{*}\partial_{\rho}h_{\sigma z*})\nonumber\\
&-4\mu^{2}(h^{+z}h_{-z}+h^{+z*}h_{-z*})+4\mu^{2}z^{*}zh^{++}h_{--}-16\mu^{2}h^{++}h_{zz*}\nonumber\\
&-2\mu^{2}(zh^{+z*}+z^{*}h^{+z})\partial_{-}h-2\mu^{2}h^{++}(z\partial_{z}h+z^{*}\partial_{z*}h)\nonumber\\
&+g^{\mu\alpha}g^{\nu\beta}g^{\rho\gamma}b_{\nu\rho}(\partial_{\mu}\partial_{\alpha}b_{\beta\gamma}-2\partial_{\mu}\partial_{\beta}b_{\alpha\gamma})+4\mu^{2}g^{\rho\gamma}b_{-\rho}(z\partial_{z}b_{-\gamma}
+z^{*}\partial_{z*}b_{-\gamma})\nonumber\\
&-4\mu^{2}g^{\rho\gamma}b_{-\rho}(z\partial_{-}b_{z\gamma}+z^{*}\partial_{-}b_{z*\gamma})
+4\mu^{2}g^{\nu\beta}b_{\nu-}(z\partial_{\beta}b_{-z}+z^{*}\partial_{\beta}b_{-z*})\nonumber\\
&-16i\mu g^{\mu\nu}\Big[h_{z*\mu}(\partial_{\nu}b_{-z}+\partial_{-}b_{z\nu}-\partial_{z}b_{\nu-})
-h_{z\mu}(\partial_{\nu}b_{-z*}+\partial_{-}b_{z*\nu}-\partial_{z*}b_{-\nu})\nonumber\\
&+h_{-\mu}(\partial_{\nu}b_{zz*}+\partial_{z}b_{z*\nu}-\partial_{z*}b_{z\nu})
-\frac{1}{2}h_{\mu\nu}(\partial_{-}b_{zz*}+\partial_{z}b_{z*-}+\partial_{z*}b_{-z})\Big]\nonumber\\
&+16g^{\mu\nu}\partial_{\mu}\phi\partial_{\nu}\phi+8g^{\mu\nu}\phi\partial_{\mu}\partial_{\nu}h
-8\phi g^{\rho\mu}g^{\nu\sigma}\partial_{\rho}\partial_{\sigma}h_{\mu\nu}-16\mu^{2}\phi h_{--}\nonumber\\
&-16\mu^{2}\phi(z\partial_{-}h_{-z}+z^{*}\partial_{-}h_{-z*})-8\mu^{2}\phi(z\partial_{z}h_{--}+z^{*}\partial_{z*}h_{--})
\nonumber\\
&-32i\mu\phi(\partial_{-}b_{zz*}+\partial_{z}b_{z*-}+\partial_{z*}b_{-z})
\Big\}.\label{FinalAction}
\end{align}
Comparing (\ref{FinalAction}) to (\ref{FinalAction2}) and replacing the coefficients: $\frac{\alpha'}{4}\leftrightarrow\frac{1}{32\kappa^{2}}$, we can confirm the correspondence between superstring field theory in the NS-NS pp-wave background and the second-order perturbation of supergravity in the same background up to total derivative. It is nontribial correspondence because the spacetime is curved and NS-NS flux exists. 


\section{Gauge symmetry in the NS-NS pp-wave}
In this section we check the gauge symmetry in the NS-NS pp-wave background at the standpoint of superstring field theory. Here we do not consider the interaction. The action is invariant in the following gauge transformation,
\begin{align}
\delta|\Phi\rangle=\hat{\delta}\cdot b^{-}_{0}|\Phi\rangle=Q_{\rm B}|\Lambda\rangle.
\end{align} 
Here the state $|\Lambda\rangle$ is a state of gauge parameter whose ghost number is one lower than the original state of string field $|\Phi\rangle$, therefore we can represent as 
\begin{align}
|\Lambda\rangle=2i\sqrt{\frac{2}{\alpha'}}\left[\hat{\epsilon}_{\mu}(x)\tilde{\beta}_{-1/2}\psi^{\mu}_{-1/2}
+\hat{\zeta}_{\mu}(x)\beta_{-1/2}\tilde{\psi}^{\mu}_{-1/2}+c^{+}_{0}\eta(x)\tilde{\beta}_{-1/2}\beta_{-1/2}\right]|\downarrow\downarrow\rangle.
\end{align}
where $\hat{\epsilon}_{\mu}(x)$, $\hat{\zeta}_{\mu}(x)$ and $\eta(x)$ are gauge parameters of the component fields.
Therefore we can obtain the gauge transformation of the component fields comparing $\delta|\Phi\rangle$ with $Q_{\rm B}|\Lambda\rangle$. The gauge transformations of $B_{\mu}$ and $E_{\mu}$ have a problem which we cannot remove the operator ordering constant $\hat{\mu}$ contained in the bosonic super-Virasoro operator, however we avoid this problem that we do not have to check the auxiliary fields $B_{\mu}$ and $E_{\mu}$ because we can integrate out these fields using the equations. Therefore ignoring these auxiliary fields and calculating the ghost parts first, $Q_{\rm B}|\Lambda\rangle$ becomes 
\begin{align}
Q_{\rm B}|\Lambda\rangle=2i\sqrt{\frac{2}{\alpha'}}&\Big[\tilde{G}^{\rm M}_{-1/2}(\hat{\epsilon}_{\mu}\psi^{\mu}_{-1/2})
+G^{\rm M}_{-1/2}(\hat{\zeta}_{\mu}\tilde{\psi}^{\mu}_{-1/2})\nonumber\\
&+\left\{\tilde{G}^{\rm M}_{1/2}(\hat{\zeta}_{\mu}\tilde{\psi}^{\mu}_{-1/2})-\eta\right\}\tilde{\gamma}_{-1/2}\beta_{-1/2}
+\left\{G^{\rm M}_{1/2}(\hat{\epsilon}_{\mu}\psi^{\mu}_{-1/2})-\eta\right\}\tilde{\beta}_{-1/2}\gamma_{-1/2}\Big]|\downarrow\downarrow\rangle.\label{GT}
\end{align}
We can calculate these terms using the formulae (\ref{der:psi:L}), (\ref{der:psi:R}), (\ref{ComRel:JF:L}) and (\ref{ComRel:JF:R}). The result is
\begin{align}
\Big[&\tilde{G}^{\rm M}_{-1/2}(\hat{\epsilon}_{\mu}\psi^{\mu}_{-1/2})
+G^{\rm M}_{-1/2}(\hat{\zeta}_{\mu}\tilde{\psi}^{\mu}_{-1/2})\Big]|\downarrow\downarrow\rangle\nonumber\\
&=\sqrt{\frac{\alpha'}{2}}\Big[-i(\partial_{\alpha}\hat{\epsilon}_{\mu}-\partial_{\mu}\hat{\zeta}_{\alpha})
\tilde{\psi}^{\alpha}_{-1/2}\psi^{\mu}_{-1/2}\nonumber\\
&+(\hat{\epsilon}_{-}-\hat{\zeta}_{-})\Big\{\frac{i}{2}\mu^{2}z^{*}\tilde{\psi}^{(+}_{-1/2}\psi^{z)}_{-1/2}
+\frac{i}{2}\mu^{2}z\tilde{\psi}^{(+}_{-1/2}\psi^{z*)}_{-1/2}
+\frac{\mu}{2}\tilde{\psi}^{[z}_{-1/2}\psi^{z*]}_{-1/2}\Big\}\nonumber\\
&+(\hat{\epsilon}_{z}-\hat{\zeta}_{z})\Big\{i\mu^{2}z\tilde{\psi}^{+}_{-1/2}\psi^{+}_{-1/2}
-\mu\tilde{\psi}^{[+}_{-1/2}\psi^{z]}_{-1/2}\Big\}
+(\hat{\epsilon}_{z*}-\hat{\zeta}_{z*})\Big\{i\mu^{2}z^{*}\tilde{\psi}^{+}_{-1/2}\psi^{+}_{-1/2}
+\mu\tilde{\psi}^{[+}_{-1/2}\psi^{z*]}_{-1/2}\Big\}\Big]|\downarrow\downarrow\rangle,
\end{align}
where the symmetric and antisymmetric symbols are defined as 
\begin{align}
\tilde{\psi}^{(\alpha}_{-1/2}\psi^{\mu)}_{-1/2}&=\tilde{\psi}^{\alpha}_{-1/2}\psi^{\mu}_{-1/2}+
\tilde{\psi}^{\mu}_{-1/2}\psi^{\alpha}_{-1/2},\\
\tilde{\psi}^{[\alpha}_{-1/2}\psi^{\mu]}_{-1/2}&=\tilde{\psi}^{\alpha}_{-1/2}\psi^{\mu}_{-1/2}-
\tilde{\psi}^{\mu}_{-1/2}\psi^{\alpha}_{-1/2}.
\end{align}
Replacing the gauge parameter with 
\begin{align}
\hat{\epsilon}_{\mu}=\zeta_{\mu}+\epsilon_{\mu}-\bar{B}_{\mu\nu}\epsilon^{\nu},\ \ \ 
\hat{\zeta}_{\mu}=\zeta_{\mu}-\epsilon_{\mu}-\bar{B}_{\mu\nu}\epsilon^{\nu},
\end{align}
and using the covariant derivative $\nabla_{\mu}$ and the background NS-NS flux $\bar{H}_{\mu\alpha\beta}$, this equation becomes 
\begin{align}
\Big[&\tilde{G}^{\rm M}_{-1/2}(\hat{\epsilon}_{\mu}\psi^{\mu}_{-1/2})
+G^{\rm M}_{-1/2}(\hat{\zeta}_{\mu}\tilde{\psi}^{\mu}_{-1/2})\Big]|\downarrow\downarrow\rangle\nonumber\\
&=-\frac{i}{2}\sqrt{\frac{\alpha'}{2}}
\Big[(\nabla_{\alpha}\epsilon_{\mu}+\nabla_{\mu}\epsilon_{\alpha})\tilde{\psi}^{(\alpha}_{-1/2}\psi^{\mu)}_{-1/2}\nonumber\\
&\ \ \ \ \ \ \ \ \ \ \ \ \ \ \ \ +\left\{\nabla_{\alpha}(\zeta_{\mu}-\bar{B}_{\mu\nu}\epsilon^{\nu})-\nabla_{\mu}(\zeta_{\alpha}-\bar{B}_{\alpha\nu}\epsilon^{\nu})+\bar{H}_{\alpha\mu\nu}\epsilon^{\nu}\right\}\tilde{\psi}^{[\alpha}_{-1/2}\psi^{\mu]}_{-1/2}\Big]|\downarrow\downarrow\rangle.\label{GT:gb}
\end{align}
Moreover we can obtain
\begin{align}
\Big[&\left\{\tilde{G}^{\rm M}_{1/2}(\hat{\zeta}_{\mu}\tilde{\psi}^{\mu}_{-1/2})-\eta\right\}
\tilde{\gamma}_{-1/2}\beta_{-1/2}
+\left\{G^{\rm M}_{1/2}(\hat{\epsilon}_{\mu}\psi^{\mu}_{-1/2})-\eta\right\}\tilde{\beta}_{-1/2}\gamma_{-1/2}\Big]|\downarrow\downarrow\rangle\nonumber\\
&=-i\sqrt{\frac{\alpha'}{2}}\Big[(g^{\alpha\mu}\partial_{\alpha}\hat{\zeta}_{\mu}-\eta)
\tilde{\gamma}_{-1/2}\beta_{-1/2}
+(g^{\alpha\mu}\partial_{\alpha}\hat{\epsilon}_{\mu}-\eta)
\tilde{\beta}_{-1/2}\gamma_{-1/2}\Big]|\downarrow\downarrow\rangle\nonumber\\
&=-\frac{i}{2}\sqrt{\frac{\alpha'}{2}}
\Big[\big\{g^{\alpha\mu}\nabla_{\alpha}(\zeta_{\mu}-\epsilon_{\mu}-\bar{B}_{\mu\nu}\epsilon^{\nu})-\eta\big\}
\tilde{\gamma}_{-1/2}\beta_{-1/2}
+\big\{g^{\alpha\mu}\nabla_{\alpha}(\zeta_{\mu}+\epsilon_{\mu}-\bar{B}_{\mu\nu}\epsilon^{\nu})-\eta\big\}
\tilde{\beta}_{-1/2}\gamma_{-1/2}\Big]|\downarrow\downarrow\rangle.\label{GT:phi}
\end{align}
where using $g^{\mu\nu}\partial_{\mu}v_{\nu}=g^{\mu\nu}\nabla_{\mu}v_{\nu}$ ($v_{\mu}$ is arbitrary covariant vector) in the NS-NS pp-wave background, we replace the partial derivative with covariant derivative in the last line of the calculation.  
Here choosing the gauge parameter $\eta$ as 
\begin{align}
\eta=g^{\alpha\mu}\nabla_{\alpha}(\zeta_{\mu}-\epsilon_{\mu}-\bar{B}_{\mu\nu}\epsilon^{\nu}),\label{eta}
\end{align}
we can eliminate the cofficient of $\tilde{\gamma}_{-1/2}\beta_{-1/2}$ namely this corresponds to gauge fixing $s(x)=0$ (see also (\ref{simplest action})).
Finally substituting (\ref{GT:gb}), (\ref{GT:phi}) and (\ref{eta}) to (\ref{GT}), we can obtain the BRST transformation of the state namely the gauge transformation as follows:
\begin{align}
Q_{\rm B}|\Lambda\rangle=\Big[&(\nabla_{\alpha}\epsilon_{\mu}+\nabla_{\mu}\epsilon_{\alpha})\tilde{\psi}^{(\alpha}_{-1/2}\psi^{\mu)}_{-1/2}\nonumber\\
&+\left\{\nabla_{\alpha}(\zeta_{\mu}-\bar{B}_{\mu\nu}\epsilon^{\nu})-\nabla_{\mu}(\zeta_{\alpha}-\bar{B}_{\alpha\nu}\epsilon^{\nu})+\bar{H}_{\alpha\mu\nu}\epsilon^{\nu}\right\}\tilde{\psi}^{[\alpha}_{-1/2}\psi^{\mu]}_{-1/2}\nonumber\\
&+2g^{\alpha\mu}\nabla_{\alpha}\epsilon_{\mu}\ \tilde{\beta}_{-1/2}\gamma_{-1/2}
+0\cdot\tilde{\gamma}_{-1/2}\beta_{-1/2}
\Big]|\downarrow\downarrow\rangle,\label{GT2}
\end{align}
where the coefficient of $\tilde{\psi}^{(\alpha}_{-1/2}\psi^{\mu)}_{-1/2}$, $\tilde{\psi}^{[\alpha}_{-1/2}\psi^{\mu]}$ and 
$\tilde{\beta}_{-1/2}\gamma_{-1/2}$, $\tilde{\gamma}_{-1/2}\beta_{-1/2}$ mean the gauge transformation of the gravitational field, the NS-NS antisymmetric field and the two scalar fields, respectively. Here we eliminate one of the two scalar fields (especially $s$) using the gauge fixing (\ref{eta}). Because of the complexity of gauge transformation for NS-NS antisymmetric tensor field, we have to explain this point.  
At the standpoint of supergravity, there are two types of the gauge transformation. One is infinitesimal transformation of general coordinates and the other is usual gauge transformation. Especially the fields with this usual gauge transformation are transformed by both gauge transformation. Here gauge transformation means both usual gauge transformation and infinitesimal transformation of general coordinates. Now such field is NS-NS antisymmetric field $B_{\mu\nu}$. Performing the infinitesimal transformation of general coordinates $x^{\mu}\rightarrow {x'}^{\mu}=x^{\mu}-\epsilon^{\mu}(x)$, metric $G_{\mu\nu}$, antisymmetric field $B_{\mu\nu}$ and dilaton $\hat{\Phi}$ are transformed as 
\begin{align}
\delta_{\epsilon}G_{\mu\nu}&=\nabla_{\mu}\epsilon_{\nu}+\nabla_{\nu}\epsilon_{\mu},\\
\delta_{\epsilon}B_{\mu\nu}&=\nabla_{\mu}(\epsilon^{\rho}B_{\rho\nu})
-\nabla_{\nu}(\epsilon^{\rho}B_{\rho\mu})+H_{\mu\nu\rho}\epsilon^{\rho},\\
\delta_{\epsilon}\hat{\Phi}&=\epsilon^{\mu}\nabla_{\mu}\hat{\Phi},
\end{align}
where we change the ordinary Lie derivative of $B_{\mu\nu}$ into the form with appearance of the flux $H_{\mu\nu\rho}$. 
 Moreover $B_{\mu\nu}$ is transformed into the following by usual gauge transformation. The result is well known as
\begin{align}
\delta_{\zeta}B_{\mu\nu}=\nabla_{\mu}\zeta_{\nu}-\nabla_{\nu}\zeta_{\mu},
\end{align}
 so that we can obtain the correct gauge transformation of $B_{\mu\nu}$ as follows:
\begin{align}
\delta B_{\mu\nu}&=\delta_{\epsilon}B_{\mu\nu}+\delta_{\zeta}B_{\mu\nu}
=\nabla_{\mu}(\zeta_{\nu}-B_{\nu\rho}\epsilon^{\rho})-\nabla_{\nu}(\zeta_{\mu}-B_{\mu\rho}\epsilon^{\rho})+H_{\mu\nu\rho}\epsilon^{\rho}.
\end{align}
Here we consider the parturbation, then the gauge transformation become
\begin{align}
\delta h_{\mu\nu}&=\nabla_{\mu}\epsilon_{\nu}+\nabla_{\nu}\epsilon_{\mu},\ \ \
(\delta h=2g^{\mu\nu}\nabla_{\mu}\epsilon_{\nu}),\label{gauge:h}\\
\delta b_{\mu\nu}&
=\nabla_{\mu}(\zeta_{\nu}-\bar{B}_{\nu\rho}\epsilon^{\rho})-\nabla_{\nu}(\zeta_{\mu}-\bar{B}_{\mu\rho}\epsilon^{\rho})+\bar{H}_{\mu\nu\rho}\epsilon^{\rho},\\
\delta{\phi}&=0\label{gauge:phi}.
\end{align}
Compared to (\ref{GT2}), we can prove the correspondence of the gauge transformation between them. Especially we just derive this nontrivial gauge transformation on $b_{\mu\nu}$ from the standpoint of superstring field theory. Moreover we have to note that the coefficient of $\tilde{\beta}_{-1/2}\gamma_{-1/2}$ in (\ref{GT}) corresponds to the gauge transformation of the trace of gravitational field $h$ because the gauge transformation of $\phi$ is zero and we can choose the scalar field arbitrarily, so that we can replace $\phi$ with $h-4\phi$ (see also section 4). Therefore as the coefficient of $\tilde{\beta}_{-1/2}\gamma_{-1/2}$, the gauge transformation of $h$ appears.    


\section{Conclusion}
In this paper, we have constructed superstring field theory in the NS-NS pp-wave background. The construction is based on the BRST first quantization of superstrings in this background. Here we have defined the NS-NS sector of the low energy state using the modes based on the {\it general operator solutions} in our previous paper. The characteristic point of the modes is coordinate dependence which plays an important role, especially it enables us to reproduce the Christoffel symbol and the covariant derivative at the standpoint of superstring field theory. Therefore we have proved the exact correspondence between the low energy string field action and the second-order perturbation of supergravity action in the same NS-NS pp-wave background. Moreover we have also proved the exact correspondence of the gauge transformation in both the actions. 

The work has a important meaning as the first step for the construction of superstring field theory in some general background fields. Before the construction, we have to construct the low energy of the other sectors namely R-R sector and R-NS (NS-R) sector. Then we can also prove the existence of supersymmetry in the background, if we prove the exact correspondence between all the sectors of superstring field action and all the terms of supergravity action in the same NS-NS pp-wave background.


\section*{Acknowledgement}
The research of S.Y. is suported in part by Rikkyo University Special Fund for Research.
\section*{Appendix}
\appendix
\section{BRST transformations of free modes in the NS-NS pp-wave}
In this section we represent the BRST transformations of free modes in the NS-NS pp-wave background. First we write up the bosonic modes.
The BRST transformations of $\tilde{\alpha}^{+}_{n},\ \alpha^{+}_{n}$ and $\tilde{\alpha}^{k}_{n},\ \alpha^{k}_{n}$ are 
\begin{align}
[Q_{\rm B},\tilde{\alpha}^{+}_{n}]=-n\sum_{m\in{\mathbb Z}}\tilde{\alpha}^{+}_{n+m}\tilde{c}_{-m}
-n\sum_{r\in{\mathbb Z}+\varepsilon}\tilde{\psi}^{+}_{n+r}\tilde{\gamma}_{-r},\ \  
[Q_{\rm B},{\alpha}^{+}_{n}]=-n\sum_{m\in{\mathbb Z}}{\alpha}^{+}_{n+m}{c}_{-m}
-n\sum_{r\in{\mathbb Z}+\varepsilon}{\psi}^{+}_{n+r}{\gamma}_{-r},\\
[Q_{\rm B},\tilde{\alpha}^{k}_{n}]=-n\sum_{m\in{\mathbb Z}}\tilde{\alpha}^{k}_{n+m}\tilde{c}_{-m}
-n\sum_{r\in{\mathbb Z}+\varepsilon}\tilde{\psi}^{+}_{n+r}\tilde{\gamma}_{-r},\ \  
[Q_{\rm B},{\alpha}^{k}_{n}]=-n\sum_{m\in{\mathbb Z}}{\alpha}^{k}_{n+m}{c}_{-m}
-n\sum_{r\in{\mathbb Z}+\varepsilon}{\psi}^{k}_{n+r}{\gamma}_{-r},
\end{align}
which are the same as flat background. The BRST transformations of $\tilde{\alpha}^{0-}_{n}$ and $\alpha^{0-}_{n}$ are
\begin{align}
[Q_{\rm B},\tilde{\alpha}^{0-}_{n}]&=-n\sum_{m\in{\mathbb Z}}\left[\tilde{\alpha}^{0-}_{n+m}-\frac{\mu}{\sqrt{2\alpha'}}\delta_{n+m,0}(J_{\rm B}+J_{\rm F})\right]\tilde{c}_{-m}-n\sum_{r\in{\mathbb Z}+\varepsilon}\tilde{\psi}^{0-}_{n+r}\tilde{\gamma}_{-r},\\
[Q_{\rm B},{\alpha}^{0-}_{n}]&=-n\sum_{m\in{\mathbb Z}}\left[{\alpha}^{0-}_{n+m}+\frac{\mu}{\sqrt{2\alpha'}}\delta_{n+m,0}(J_{\rm B}+J_{\rm F})\right]{c}_{-m}-n\sum_{r\in{\mathbb Z}+\varepsilon}{\psi}^{0-}_{n+r}{\gamma}_{-r}.
\end{align}
Here we have to note that the BRST transformations of $\tilde{\alpha}^{0-}_{n}$ and $\alpha^{0-}_{n}$ contain both left modes and right modes in $J_{\rm B}$ and $J_{\rm F}$ so that we cannot divide them into left and right. 
The BRST transformations of $\hat{A}_{n},\ \hat{A}^{\dagger}_{n}$ and $\hat{B}_{n},\ \hat{B}^{\dagger}_{n}$ are 
\begin{align}
[Q_{\rm B},\hat{A}_{n}]=&\sum_{m\in{\mathbb Z}}\left[-\mu\sqrt{\frac{\alpha'}{2}}\tilde{\alpha}^{+}_{m}\hat{A}_{n}
-(n+m-\hat{\mu})\hat{A}_{n+m}\right]\tilde{c}_{-m}+\sum_{m\in{\mathbb Z}}\mu\sqrt{\frac{\alpha'}{2}}{\alpha}^{+}_{m}\hat{A}_{n}c_{-m}\nonumber\\
+&\sum_{r\in{\mathbb Z}+\varepsilon}\left[-\mu\sqrt{\frac{\alpha'}{2}}\tilde{\psi}^{+}_{r}\hat{A}_{n}-i\tilde{\lambda}_{r+n}\right]\tilde{\gamma}_{-r}+\sum_{r'\in{\mathbb Z}+\varepsilon}\mu\sqrt{\frac{\alpha'}{2}}{\psi}^{+}_{r'}\hat{A}_{n}
\gamma_{-r'},\\
[Q_{\rm B},\hat{A}^{\dagger}_{n}]=&\sum_{m\in{\mathbb Z}}\left[\mu\sqrt{\frac{\alpha'}{2}}\tilde{\alpha}^{+}_{m}\hat{A}^{\dagger}_{n}
+(n-m-\hat{\mu})\hat{A}^{\dagger}_{n-m}\right]\tilde{c}_{-m}-\sum_{m\in{\mathbb Z}}\mu\sqrt{\frac{\alpha'}{2}}{\alpha}^{+}_{m}\hat{A}^{\dagger}_{n}c_{-m}\nonumber\\
+&\sum_{r\in{\mathbb Z}+\varepsilon}\left[\mu\sqrt{\frac{\alpha'}{2}}\tilde{\psi}^{+}_{r}\hat{A}^{\dagger}_{n}-i\tilde{\lambda}^{\dagger}_{-r+n}\right]\tilde{\gamma}_{-r}-\sum_{r'\in{\mathbb Z}+\varepsilon}\mu\sqrt{\frac{\alpha'}{2}}{\psi}^{+}_{r'}\hat{A}^{\dagger}_{n}
\gamma_{-r'},\\
[Q_{\rm B},\hat{B}_{n}]=&\sum_{m\in{\mathbb Z}}\left[\mu\sqrt{\frac{\alpha'}{2}}{\alpha}^{+}_{m}\hat{B}_{n}
-(n+m+\hat{\mu})\hat{B}_{n+m}\right]{c}_{-m}-\sum_{m\in{\mathbb Z}}\mu\sqrt{\frac{\alpha'}{2}}\tilde{\alpha}^{+}_{m}\hat{B}_{n}\tilde{c}_{-m}\nonumber\\
+&\sum_{r\in{\mathbb Z}+\varepsilon}\left[\mu\sqrt{\frac{\alpha'}{2}}{\psi}^{+}_{r}\hat{B}_{n}-i{\lambda}_{r+n}\right]
{\gamma}_{-r}-\sum_{r'\in{\mathbb Z}+\varepsilon}\mu\sqrt{\frac{\alpha'}{2}}\tilde{\psi}^{+}_{r'}\hat{B}_{n}
\tilde{\gamma}_{-r'},\\
[Q_{\rm B},\hat{B}^{\dagger}_{n}]=&\sum_{m\in{\mathbb Z}}\left[-\mu\sqrt{\frac{\alpha'}{2}}{\alpha}^{+}_{m}
\hat{B}^{\dagger}_{n}
+(n-m+\hat{\mu})\hat{B}^{\dagger}_{n-m}\right]{c}_{-m}+\sum_{m\in{\mathbb Z}}\mu\sqrt{\frac{\alpha'}{2}}
\tilde{\alpha}^{+}_{m}\hat{B}^{\dagger}_{n}\tilde{c}_{-m}\nonumber\\
+&\sum_{r\in{\mathbb Z}+\varepsilon}\left[-\mu\sqrt{\frac{\alpha'}{2}}{\psi}^{+}_{r}\hat{B}^{\dagger}_{n}-i{\lambda}^{\dagger}_{-r+n}\right]{\gamma}_{-r}+\sum_{r'\in{\mathbb Z}+\varepsilon}\mu\sqrt{\frac{\alpha'}{2}}\tilde{\psi}^{+}_{r'}\hat{B}^{\dagger}_{n}
\gamma_{-r'},
\end{align}
which also contain both left and right modes.
Here we distinguish the sum indeces $r$ and $r'$ because of the possibility that left modes and right modes belong to the different sector.
Next we write up the fermionic modes. The BRST transformations of $\tilde{\psi}^{+}_{s},\ \psi^{+}_{s}$ and $\tilde{\psi}^{k}_{s},\ \psi^{k}_{s}$ are 
\begin{align}
\{Q_{\rm B},\tilde{\psi}^{+}_{s}\}=\sum_{m\in{\mathbb Z}}(s+\frac{m}{2})\tilde{\psi}^{+}_{s+m}\tilde{c}_{-m}
+\sum_{r\in{\mathbb Z}+\varepsilon}\tilde{\alpha}^{+}_{s+r}\tilde{\gamma}_{-r},\ 
\{Q_{\rm B},{\psi}^{+}_{s}\}=\sum_{m\in{\mathbb Z}}(s+\frac{m}{2}){\psi}^{+}_{s+m}{c}_{-m}
+\sum_{r\in{\mathbb Z}+\varepsilon}{\alpha}^{+}_{s+r}{\gamma}_{-r},\\
\{Q_{\rm B},\tilde{\psi}^{k}_{s}\}=\sum_{m\in{\mathbb Z}}(s+\frac{m}{2})\tilde{\psi}^{k}_{s+m}\tilde{c}_{-m}
+\sum_{r\in{\mathbb Z}+\varepsilon}\tilde{\alpha}^{k}_{s+r}\tilde{\gamma}_{-r},\ 
\{Q_{\rm B},{\psi}^{k}_{s}\}=\sum_{m\in{\mathbb Z}}(s+\frac{m}{2}){\psi}^{k}_{s+m}{c}_{-m}
+\sum_{r\in{\mathbb Z}+\varepsilon}{\alpha}^{k}_{s+r}{\gamma}_{-r}.
\end{align}
which are the same as flat background.
The BRST transformation of $\tilde{\psi}^{0-}_{s}$ and $\psi^{0-}_{s}$ are
\begin{align}
\{Q_{\rm B},\tilde{\psi}^{0-}_{s}\}&=\sum_{m\in{\mathbb Z}}(s+\frac{m}{2})\tilde{\psi}^{0-}_{s+m}\tilde{c}_{-m}
+\sum_{r\in{\mathbb Z}+\varepsilon}\left[\tilde{\alpha}^{0-}_{s+r}-\delta_{s+r,0}\frac{\mu}{\sqrt{2\alpha'}}
(J_{\rm B}+J_{\rm F})\right]\tilde{\gamma}_{-r},\\
\{Q_{\rm B},{\psi}^{0-}_{s}\}&=\sum_{m\in{\mathbb Z}}(s+\frac{m}{2}){\psi}^{0-}_{s+m}{c}_{-m}
+\sum_{r\in{\mathbb Z}+\varepsilon}\left[{\alpha}^{0-}_{s+r}+\delta_{s+r,0}\frac{\mu}{\sqrt{2\alpha'}}
(J_{\rm B}+J_{\rm F})\right]{\gamma}_{-r}.
\end{align}
Finally the BRST transformations of $\tilde{\lambda}_{s}$, $\tilde{\lambda}^{\dagger}_{s}$ and $\lambda_{s}$, $\lambda^{\dagger}_{s}$ are
\begin{align}
\{Q_{\rm B},\tilde{\lambda}_{s}\}=&\sum_{m\in{\mathbb Z}}
\left[\mu\sqrt{\frac{\alpha'}{2}}\tilde{\alpha}^{+}_{m}\tilde{\lambda}_{s}+(s+\frac{m}{2}-\hat{\mu})\tilde{\lambda}_{s+m}\right]\tilde{c}_{-m}
-\sum_{m\in{\mathbb Z}}\mu\sqrt{\frac{\alpha'}{2}}\alpha^{+}_{m}\tilde{\lambda}_{s}c_{-m}\nonumber\\
&-\sum_{r\in{\mathbb Z}+\varepsilon}\left[\mu\sqrt{\frac{\alpha'}{2}}\tilde{\psi}^{+}_{r}\tilde{\lambda}_{s}
+i(s+r-\hat{\mu})\hat{A}_{s+r}\right]\tilde{\gamma}_{-r}
+\sum_{r'\in{\mathbb Z}+\varepsilon}\mu\sqrt{\frac{\alpha'}{2}}\psi^{+}_{r'}\tilde{\lambda}_{s}\gamma_{-r'},\\
\{Q_{\rm B},\tilde{\lambda}_{s}^{\dagger}\}=&-\sum_{m\in{\mathbb Z}}\left[\mu\sqrt{\frac{\alpha'}{2}}\tilde{\alpha}^{+}_{m}\tilde{\lambda}^{\dagger}_{s}
+(s-\frac{m}{2}-\hat{\mu})\tilde{\lambda}^{\dagger}_{s-m}\right]\tilde{c}_{-m}
+\sum_{m\in{\mathbb Z}}\mu\sqrt{\frac{\alpha'}{2}}\alpha^{+}_{m}\tilde{\lambda}^{\dagger}_{s}c_{-m}\nonumber\\
&+\sum_{r\in{\mathbb Z}+\varepsilon}\left[\mu\sqrt{\frac{\alpha'}{2}}\tilde{\psi}^{+}_{r}\tilde{\lambda}^{\dagger}_{s}
+i(s-r-\hat{\mu})\hat{A}^{\dagger}_{s-r}\right]\tilde{\gamma}_{-r}
-\sum_{r'\in{\mathbb Z}+\varepsilon}\mu\sqrt{\frac{\alpha'}{2}}\psi^{+}_{r'}\tilde{\lambda}^{\dagger}_{s}\gamma_{-r'},\\
\{Q_{\rm B},\lambda_{s}\}=&-\sum_{m\in{\mathbb Z}}
\left[\mu\sqrt{\frac{\alpha'}{2}}\alpha^{+}_{m}\lambda_{s}-(s+\frac{m}{2}+\hat{\mu})\lambda_{s+m}\right]c_{-m}
+\sum_{m\in{\mathbb Z}}\mu\sqrt{\frac{\alpha'}{2}}\tilde{\alpha}^{+}_{m}\lambda_{s}\tilde{c}_{-m}\nonumber\\
&+\sum_{r\in{\mathbb Z}+\varepsilon}\left[\mu\sqrt{\frac{\alpha'}{2}}\psi^{+}_{r}\lambda_{s}-i(s+r+\hat{\mu})\hat{B}_{s+r}\right]\gamma_{-r}-\sum_{r'\in{\mathbb Z}+\varepsilon}\mu\sqrt{\frac{\alpha'}{2}}\tilde{\psi}^{+}_{r'}\lambda_{s}
\tilde{\gamma}_{-r'},\\
\{Q_{\rm B},\lambda^{\dagger}_{s}\}=&\sum_{m\in{\mathbb Z}}
\left[\mu\sqrt{\frac{\alpha'}{2}}\alpha^{+}_{m}\lambda^{\dagger}_{s}-(s-\frac{m}{2}+\hat{\mu})\lambda^{\dagger}_{s-m}\right]c_{-m}-\sum_{m\in{\mathbb Z}}\mu\sqrt{\frac{\alpha'}{2}}\tilde{\alpha}^{+}_{m}\lambda^{\dagger}_{s}\tilde{c}_{-m}
\nonumber\\
&-\sum_{r\in{\mathbb Z}+\varepsilon}\left[\mu\sqrt{\frac{\alpha'}{2}}\psi^{+}_{r}\lambda^{\dagger}_{s}
-i(s-r+\hat{\mu})\hat{B}^{\dagger}_{s-r}\right]\gamma_{-r}
+\sum_{r'\in{\mathbb Z}+\varepsilon}\mu\sqrt{\frac{\alpha'}{2}}\tilde{\psi}^{+}_{r'}\lambda^{\dagger}_{s}\tilde{\gamma}_{-r'}.
\end{align}
The BRST transformations of free modes in NS-NS pp-wave background contain interaction between left modes and right modes. The BRST transformations of ghost modes are the same as the flat background. 
\section{Definition of Modes}
In this section we define the modes based on the {\it general operator solutions} and the {\it free mode representations}. First of all, we have to define coordinates of the center of mass $z,\ z^{*}$ and total momenta $p_{z},\ p_{z*}$. Here we define them as 
\begin{align}
z&=\int_{0}^{2\pi}\frac{d\sigma}{2\pi}Z(\tau,\sigma)|_{\tau=0},\ \ \ \ 
z^{*}=\int_{0}^{2\pi}\frac{d\sigma}{2\pi}Z^{*}(\tau,\sigma)|_{\tau=0},\\
p_{z}&=\int_{0}^{2\pi}d\sigma P_{Z}(\tau,\sigma)|_{\tau=0},\ \ p_{z*}=\int_{0}^{2\pi}d\sigma P_{Z*}(\tau,\sigma)|_{\tau=0},
\end{align}
where we defined $P_{Z},\ P_{Z*}$ in our previous paper as 
\begin{align}
P_{Z}&=\frac{1}{4\pi\alpha'}e^{i\mu\tilde{X}^{+}}\left[\partial_{+}f^{*}+\partial_{-}g^{*}+\mu(\psi^{+}_{+}\lambda^{*}_{+}-
\psi^{+}_{-}\lambda^{*}_{-})\right],\\
P_{Z*}&=\frac{1}{4\pi\alpha'}e^{-i\mu\tilde{X}^{+}}\left[\partial_{+}f+\partial_{-}g-\mu(\psi^{+}_{+}\lambda_{+}-
\psi^{+}_{-}\lambda_{-})\right].
\end{align}
Here we construct $z$ representatively. Now we consider the case of $\tau=0$, so that $z$ becomes  
\begin{align}
z=\int_{0}^{2\pi}\frac{d\sigma}{2\pi}e^{-i\mu\tilde{X}^{+}(\sigma)}\left[f(\sigma)+g(-\sigma)\right].
\end{align} 
Then we have to consider the factor $e^{-i\mu\tilde{X}^{+}(\sigma)}$. Replacing $n\rightarrow -n$ to the right moving mode, in the case of $\tau=0$, $\tilde{X}^{+}(\sigma)$ becomes
\begin{align}
\tilde{X}^{+}(\sigma)=\alpha'p^{+}\sigma+i\sqrt{\frac{\alpha'}{2}}\sum_{n\neq 0}\frac{1}{n}(\tilde{\alpha}^{+}
+\alpha^{+}_{-n})e^{-in\sigma},
\end{align}
so that the factor $e^{-i\mu\tilde{X}^{+}(\sigma)}$ becomes
\begin{align}
e^{-i\mu\tilde{X}^{+}(\sigma)}=e^{-i\hat{\mu}\sigma}\prod_{n\neq 0}\exp\left[\mu\sqrt{\frac{\alpha'}{2}}\frac{1}{n}(\tilde{\alpha}^{+}
+\alpha^{+}_{-n})e^{-in\sigma}\right].
\end{align}
Here we define $\Xi(\sigma)$ for convinience as 
\begin{align}
\Xi(\sigma)&=\prod_{n\neq 0}\exp\left[\mu\sqrt{\frac{\alpha'}{2}}\frac{1}{n}(\tilde{\alpha}^{+}
+\alpha^{+}_{-n})e^{-in\sigma}\right]\nonumber\\
&=\prod_{n\neq 0}\sum_{k=0}^{\infty}\frac{1}{k!}\left(\mu\sqrt{\frac{\alpha'}{2}}\right)^{k}\left[\frac{1}{n}(\tilde{\alpha}^{+}
+\alpha^{+}_{-n})\right]^{k}e^{-ink\sigma},
\end{align}
where we expand $\Xi(\sigma)$ in the last line of this equation.
Here we also define 
\begin{align}
\xi_{n,k}=\frac{1}{k!}\left(\mu\sqrt{\frac{\alpha'}{2}}\right)^{k}\left[\frac{1}{n}(\tilde{\alpha}^{+}
+\alpha^{+}_{-n})\right]^{k},
\end{align}
Here we have to note that $\xi_{n,k=0}=1$ is very important property.
Therefore we can rewrite the factor $e^{-i\mu\tilde{X}^{+}}$ as
\begin{align}
e^{-i\mu\tilde{X}^{+}}=e^{-i\hat{\mu}\sigma}\Xi(\sigma)=\prod_{n\neq 0}\sum_{k=0}^{\infty}\xi_{k,n}e^{-i(\hat{\mu}+nk)\sigma}.   
\end{align}
So that $z$ becomes
\begin{align}
z=\int_{0}^{2\pi}\frac{d\sigma}{2\pi}e^{-i\hat{\mu}\sigma}\Xi(\sigma)\left[f(\sigma)+g(-\sigma)\right].
\end{align}
Substituting the free mode representations of $f(\sigma)$ and $g(-\sigma)$, the factor $\hat{\mu}$ vanish and finally we can obtain the following:
\begin{align}
z&=\sqrt{\alpha'}\prod_{n\neq 0}\sum_{k=0}^{\infty}\xi_{n,k}(\hat{A}_{-nk}+\hat{B}_{nk})\nonumber\\
&=\sqrt{\alpha'}(\hat{A}_{0}+\hat{B}_{0})+\sqrt{\alpha'}\prod_{n\neq 0}\sum_{k>0}\xi_{n,k}(\hat{A}_{-nk}+\hat{B}_{nk}).
\end{align}
Here we can obtain $z^{*}$ taking Hermitian conjugate of $z$. Now we consider the state of the coordinate representation $|x\rangle$. Taking the expectation value $\langle x|z|x\rangle$, in this equation, the second term vanish so that in substance we can define $z$ and $z^{*}$ as follows:
\begin{align}
z=\sqrt{\alpha'}(\hat{A}_{0}+\hat{B}_{0}),\ \ 
z^{*}=\sqrt{\alpha'}(\hat{A}^{\dagger}_{0}+\hat{B}^{\dagger}_{0}).
\label{COM:zz*}
\end{align} 
Similarly substituting the free mode representations, we can obtain $p_{z*}$:
\begin{align}
p_{z*}=\frac{i}{2\sqrt{\alpha'}}\hat{\mu}(\hat{A}_{0}-\hat{B}_{0})+\frac{1}{2}\prod_{n\neq 0}\sum_{k>0}\xi_{n,k}
\left[\frac{i}{\sqrt{\alpha'}}(nk+\hat{\mu})(\hat{A}_{-nk}+\hat{B}_{nk})-\sqrt{2}\mu
\sum_{r}(\tilde{\psi}^{+}_{r}\tilde{\lambda}_{-r-nk}-\psi^{+}_{-r}\lambda_{r+nk})\right].
\end{align}
Taking expectation value $\langle x|p_{z*}|x\rangle$, the second term vanish so that we can also define 
\begin{align}
p_{z*}=\frac{i}{2\sqrt{\alpha'}}\hat{\mu}(\hat{A}_{0}-\hat{B}_{0}),\ \ 
p_{z}=-\frac{i}{2\sqrt{\alpha'}}\hat{\mu}(\hat{A}^{\dagger}_{0}-\hat{B}^{\dagger}_{0}),
\end{align}
in substance.
Therefore we can write $\hat{A}_{0}$ and $\hat{B}_{0}$ using $z$ and $p_{z*}$ as follows:
\begin{align}
\hat{A}_{0}=\frac{1}{2\sqrt{\alpha'}}(z-2i\alpha'\hat{\mu}^{-1}p_{z*}),\ \ 
\hat{B}_{0}=\frac{1}{2\sqrt{\alpha'}}(z+2i\alpha'\hat{\mu}^{-1}p_{z*}),
\label{A0B0}
\end{align}
so that these modes behave as the harmonic oscillator. $\hat{A}_{0}^{\dagger}$ and $\hat{B}_{0}^{\dagger}$ are defined by taking the Hermitian conjugate of them.
Next we define the modes $\tilde{\psi}^{z}_{-1/2}$, $\psi^{z}_{-1/2}$, their Hermitian conjugate and $\tilde{\psi}^{-}_{-1/2}$, $\psi^{-}_{-1/2}$. First we can define the modes generally as follows because of the (anti)periodicity of $\psi^{\mu}_{\pm}(\tau,\sigma)$.
\begin{align}
\tilde{\psi}^{\mu}_{r}&=\frac{1}{\sqrt{\alpha'}}\int_{0}^{2\pi}\frac{d\sigma}{2\pi}e^{ir\sigma}\psi^{\mu}_{+}(\sigma^{+},\sigma^{-})|_{\tau=0},\ \ 
\psi^{\mu}_{r}=\frac{1}{\sqrt{\alpha'}}\int_{0}^{2\pi}\frac{d\sigma}{2\pi}e^{-ir\sigma}\psi^{\mu}_{-}(\sigma^{+},\sigma^{-})|_{\tau=0},
\end{align}
where $r\in\mathbb{Z}+\varepsilon$ where $\varepsilon=0$ for Ramond sector and $\varepsilon=1/2$ for Neveu-Schwarz sector, and we consider $\tau=0$. Here we consider $\tilde{\psi}^{z}_{s}$ representatively. Substituting the {\it free-mode representations}, $\tilde{\psi}^{z}_{s}$ becomes  
\begin{align}
\tilde{\psi}^{z}_{s}=\int_{0}^{2\pi}\frac{d\sigma}{2\pi}\prod_{m\neq 0}\sum_{k=0}^{\infty}\xi_{m,k}
\left[\sqrt{2}\sum_{r}\tilde{\lambda}_{r}e^{i(s-r-mk)\sigma}
-i\mu\sqrt{\alpha'}\sum_{r,n}\tilde{\psi}^{+}_{r}(\hat{A}_{n}+\hat{B}_{-n})
e^{i(s-r-n-mk)\sigma}\right].
\end{align}
Calculating the integral we can make the Kronecker's delta and taking the summation over $r$ the equation becomes 
\begin{align}
\tilde{\psi}^{z}_{s}=\prod_{m\neq 0}\sum_{k=0}\xi_{m,k}\left[
\sqrt{2}\tilde{\lambda}_{s-mk}-i\mu\sqrt{\alpha'}
\sum_{n}\tilde{\psi}^{+}_{s-n-mk}(\hat{A}_{n}+\hat{B}_{-n})
\right].
\end{align}
Here only in the case of $k=0$ and $n=0$, the state created by $\tilde{\psi}^{z}_{s<0}$ can become the eigenstate of Hamiltonian $H=\tilde{L}_{0}+L_{0}$ whose energy eigenvalue is $(-s)$ namely $H\tilde{\psi}^{z}_{s}|0\rangle=-s\tilde{\psi}^{z}_{s}|0\rangle$ therefore we can practically define $\tilde{\psi}^{z}_{s}$ as 
\begin{align}
\tilde{\psi}^{z}_{s}=\sqrt{2}\tilde{\lambda}_{s}-i\mu\sqrt{\alpha'}\tilde{\psi}^{+}_{s}(\hat{A}_{0}+\hat{B}_{0}).
\end{align} 
Similarly substituting the {\it free-mode representations}, $\tilde{\psi}^{-}_{s}$ becomes
\begin{align}
\tilde{\psi}^{-}_{s}=\int\frac{d\sigma}{2\pi}
\left[\sum_{r}\tilde{\psi}^{0-}_{r}e^{i(s-r)\sigma}-i\mu\sqrt{\frac{\alpha'}{2}}\sum_{r,n}
\left\{\tilde{\lambda}^{\dagger}_{r}(\hat{A}_{n}+\hat{B}_{-n})e^{i(s+r-n)\sigma}-\tilde{\lambda}_{r}
(\hat{A}^{\dagger}_{n}+\hat{B}^{\dagger}_{-n})e^{i(s-r+n)\sigma}\right\}\right].
\end{align}
Here we make Kroncker's delta symbol by integration over $\sigma$, and carrying out the summation over $r$, we can obtain 
\begin{align}
\tilde{\psi}^{-}_{s}=\tilde{\psi}^{0-}_{s}-i\mu\sqrt{\frac{\alpha'}{2}}\sum_{n}
\left[\tilde{\lambda}^{\dagger}_{n-s}(\hat{A}_{n}+\hat{B}_{-n})
-\tilde{\lambda}_{n+s}(\hat{A}^{\dagger}_{n}+\hat{B}^{\dagger}_{-n})\right].
\end{align}
Then only the case of $n=0$ mode can become eigenstate of Hamiltonian, therefore we can define practially
\begin{align}
\tilde{\psi}^{-}_{s}=\tilde{\psi}^{0-}_{s}-i\mu\sqrt{\frac{\alpha'}{2}}
\left[\tilde{\lambda}^{\dagger}_{-s}(\hat{A}_{0}+\hat{B}_{0})
-\tilde{\lambda}_{s}(\hat{A}^{\dagger}_{0}+\hat{B}^{\dagger}_{0})\right].
\end{align} 
Finally using the coodinates $z$ and $z^{*}$ defined in (\ref{COM:zz*}), we can obtain the following equations:
\begin{align}
\tilde{\psi}^{z}_{s}&=\sqrt{2}\tilde{\lambda}_{s}-i\mu z\tilde{\psi}^{+}_{s},\ \ \ \ \ \ \ \ \ \ \ \ \ \ \ {\psi}^{z}_{s}=\sqrt{2}{\lambda}_{s}+i\mu z{\psi}^{+}_{s},\label{psi^z_s}\\
\tilde{\psi}^{-}_{s}&=\tilde{\psi}^{0-}_{s}-\frac{i\mu}{\sqrt{2}}[z\tilde{\lambda}^{\dagger}_{-s}-z^{*}\tilde{\lambda}_{s}],\ \ 
{\psi}^{-}_{s}={\psi}^{0-}_{s}+\frac{i\mu}{\sqrt{2}}[z{\lambda}^{\dagger}_{-s}-z^{*}{\lambda}_{s}].
\label{psi^-_s}
\end{align}
These modes play an important role for construction of the low energy string field action.


\end{document}